\documentclass[sigconf,twocolumn, natbib=false, anonymous=false]{acmart}

\AtBeginDocument{%
  \providecommand\BibTeX{{%
    Bib\TeX}}}

\setcopyright{none}
\acmISBN{}
\acmDOI{10.48550/arXiv.2507.05289}
\acmConference{Submitted for review}{May 25}
  
\usepackage{amsmath,amsfonts}
\usepackage{algorithmic}
\usepackage{graphicx}
\usepackage{hyperref}
\usepackage{textcomp}
\usepackage[backend=biber, style=numeric, datamodel=acmdatamodel, sorting=none]{biblatex}
\usepackage{multirow}
\usepackage{tabularx} 
\usepackage{caption}
\usepackage{xcolor}
\usepackage{booktabs}
\usepackage{amsmath}
\usepackage{wrapfig}
\usepackage{acronym}
\usepackage{enumitem}
\usepackage{tablefootnote}
\usepackage{listings}
\usepackage{tcolorbox} 
\usepackage{colortbl} 
\usepackage{fancyvrb} 

\def\BibTeX{{\rm B\kern-.05em{\sc i\kern-.025em b}\kern-.08em
    T\kern-.1667em\lower.7ex\hbox{E}\kern-.125emX}}

\definecolor{backcolour}{rgb}{1,1,1} 
\definecolor{codegreen}{rgb}{0,0.6,0}
\definecolor{codegray}{rgb}{0.5,0.5,0.5}
\definecolor{codepurple}{rgb}{0.58,0,0.82}
\definecolor{mymauve}{rgb}{0.58,0,0.82}
\definecolor{mygreen}{rgb}{0,0.6,0}

\lstdefinestyle{jsonstyle}{
    backgroundcolor=\color{backcolour},   
    commentstyle=\color{codegreen},
    keywordstyle=\color{magenta},
    numberstyle=\tiny\color{codegray},
    stringstyle=\color{codepurple},
    basicstyle=\ttfamily\footnotesize,
    breakatwhitespace=false,         
    breaklines=true,                 
    captionpos=b,                    
    keepspaces=true,                 
    numbers=left,                    
    numbersep=5pt,                  
    showspaces=false,                
    showstringspaces=false,
    showtabs=false,                  
    tabsize=2
}

\lstdefinelanguage{json}{
    basicstyle=\normalfont\ttfamily,
    numbers=left,
    numberstyle=\scriptsize,
    basicstyle=\ttfamily\footnotesize,
    stepnumber=1,
    numbersep=8pt,
    showstringspaces=false,
    breaklines=true,
    frame=lines,
    backgroundcolor=\color{backcolour},
    stringstyle=\color{codegreen},
    keywordstyle=\color{mymauve},
    commentstyle=\color{mygreen},
    escapeinside={(*@}{@*)},
    morestring=[b]",
    morestring=[d]'
}

\lstdefinestyle{promptstyle}{
    backgroundcolor=\color{backcolour},   
    basicstyle=\ttfamily\footnotesize,
    breakatwhitespace=true,         
    breaklines=true,                 
    captionpos=b,                    
    keepspaces=true,                 
    numbers=none,                    
    numbersep=5pt,                  
    showspaces=false,                
    showstringspaces=false,
    showtabs=false,                  
    tabsize=0
}

\lstdefinelanguage{prompt}{
    basicstyle=\normalfont\ttfamily,
    numbers=none,
    numberstyle=\scriptsize,
    stepnumber=1,
    numbersep=8pt,
    showstringspaces=false,
    breaklines=true,
    breakindent=0pt,
    frame=lines,
    backgroundcolor=\color{backcolour},
    escapeinside={(*@}{@*)},
    morestring=[b]",
    morestring=[d]'
}

\newacro{LLM}[LLM]{Large Language Model}
\newacro{FSL}[FSL]{Few-Shot Learning}
\newacro{ZSL}[ZSL]{Zero-Shot Learning}
\newacro{AST}[AST]{Abstract Syntax Tree}
\newacro{CNN}[CNN]{Convolutional Neural Network}
\newacro{OSS}[OSS]{Opens Source Software}
\newacro{SC}[SC]{Scalabrino Classifier}

\begin{document}
\title{Measuring how changes in code readability attributes affect code quality evaluation by Large Language Models}

\author{Igor Regis da Silva Simões}
\email{igor@bb.com.br}
\orcid{0009-0003-8129-3297}
\affiliation{
  \institution{Banco do Brasil / Universidade de Brasília}
  \city{Brasília}
  \country{Brazil}}

\author{Elaine Venson}
\email{elainevenson@unb.br}
\affiliation{%
  \institution{Universidade de Brasília}
  \city{Brasília}
  \state{DF}
  \country{Brazil}
}

\begin{abstract}
Code readability is one of the main aspects of code quality, influenced by various properties like identifier names, comments, code structure, and adherence to standards. 
However, measuring this attribute poses challenges in both industry and academia. While static analysis tools assess attributes such as code smells and comment percentage, code reviews introduce an element of subjectivity.
This paper explores using Large Language Models (LLMs) to evaluate code quality attributes related to its readability in a standardized, reproducible, and consistent manner. 
We conducted a quasi-experiment study to measure the effects of code changes on \ac{LLM}s' interpretation regarding its readability quality attribute. Nine LLMs were tested, undergoing three interventions: removing comments, replacing identifier names with obscure names, and refactoring to remove code smells. Each intervention involved 10 batch analyses per LLM, collecting data on response variability. We compared the results with a known reference model and tool.
The results showed that all LLMs were sensitive to the interventions, with agreement with the reference classifier being high for the original and refactored code scenarios. However, this agreement diverged for the other two interventions. The LLMs demonstrated a strong semantic sensitivity that the reference model did not fully capture. A thematic analysis of the LLMs' reasoning confirmed their evaluations directly reflected the nature of each intervention. The models also exhibited response variability, with 9.37\% to 14.58\% of executions showing a standard deviation greater than zero, indicating response oscillation, though this did not always compromise the statistical significance of the results.
LLMs demonstrated potential for evaluating semantic quality aspects, such as coherence between identifier names, comments, and documentation with code purpose. Further research is needed to compare these evaluations with human assessments and explore real-world application limitations, including cost factors.
\end{abstract}

\keywords{Code Quality, Code Comprehensibility, Static Analysis, Software Engineering, LLM, ChatGPT, Gemini, Llama, Claude}

\maketitle

\captionsetup{font=footnotesize}

\section{Introduction}

\newcommand{\gptO}{ChatGPT 4o}
\newcommand{\gptMini}{ChatGPT 4o-mini}
\newcommand{\geminiPro}{Gemini 2.0 Pro}
\newcommand{\geminiFlash}{Gemini 2.0 Flash}
\newcommand{\llamaQuatrocentos}{Llama 3.1-405B}
\newcommand{\llamaOito}{Llama 3.1-8B}
\newcommand{\claudeSonnet}{Claude 3.7 Sonnet}
\newcommand{\claudeHaiku}{Claude 3.5 Haiku}
\newcommand{\deepSeek}{Deep Seek V3}
\setlength{\tabcolsep}{2.5pt}

Reading source code is necessary for developers to understand the software's purpose, making it a central activity in software maintenance \cite{raymond_reading_1991}. This activity of understanding the software or program can occur top-down or bottom-up. In the top-down approach, the developer starts by reading documents such as requirements and architectural documents, then moves to implementation layers. In the bottom-up approach, the developer starts directly from the code implementation \cite{obrien_expectation_2004}.


Source code can be readable and comprehensible. These two characteristics are often treated as the same, but they are distinct textual features. Readability relates to the complexity of the presented text, while comprehensibility relates to the reader's ability to grasp the text's meaning. Thus, code measured as readable may still be incomprehensible to a reader lacking the necessary knowledge or experience to understand it \cite{boehm_quantitative_1976}.

This research focuses on two commonly studied readability attributes of source code - identifier names and code comments, as detailed in Section \ref{sec:mehodology}. We do not address other artifacts that comprise the software (documentation, architecture, etc.). Also, our study do not cover the code understandability research area, as it would require evaluating developers and their characteristics, such as seniority and familiarity with the language in which the code is written \cite{scalabrino_automatically_2019}.

Low code readability is treated as a type of technical debt. Static code analysis tools aim to map violations of coding best practices to measure this debt. SonarQube classifies this type of violation as: \textit{``Code Debt - Refers to problems found in the source code (violating best practices or coding rules) that negatively affect its readability and make it difficult to maintain.''}\footnote{SonarQube https://www.sonarsource.com/learn/technical-debt/\#types-of-technical-debt} \cite{kruchten_managing_2019}. 

Low source code readability affects the industry, causing negative impacts over developers' productivity \cite{cheng_what_2022} and job satisfaction \cite{besker_influence_2020}. Many research studies have been conducted to advance the understanding of code readability metrics \cite{wyrich_40_2024}. Several approaches have been developed, yet none of them reach the same level of accordance with human evaluation, showing that there is still gaps to be filled \cite{buse_metric_2008, dorn_general_2012, scalabrino_improving_2016, scalabrino_comprehensive_2018, mi_inception_2018, mi_towards_2022}. 

The rise of \ac{LLM}s and their emergent abilities, opens up new avenues for research. There are works investigating which code characteristics LLMs are sensitive to \cite{hu_how_2024}, how LLMs can be used to evaluate code quality \cite{simoes_evaluating_2024}, and how they can support personalized assessments of code readability for individual developers \cite{vitale_personalized_2025}. These preliminary results point to the need for a better understanding of how \ac{LLM} assess code readability.

In this work, we explore these issues, addressing the following research questions:

\newcommand{\perguntaUm}{Can LLMs be used to generate a metric that assesses code quality attributes associated with readability?}
\newcommand{\perguntaDois}{How code attributes such as identifier naming and code comments affect readability and quality perception by an LLM?}
\newcommand{\perguntaTres}{How can the inherent variability of LLMs affect their evaluation of code readability and quality?}
\begin{enumerate}
    \item \textbf{\perguntaUm}\label{pergunta1}
    \item \textbf{\perguntaDois}\label{pergunta2}
    \item \textbf{\perguntaTres}\label{pergunta3}
\end{enumerate}

The remainder of this paper is organized as follows: 
Section \ref{sec:relatedWork}, explores the studies addressing the researched topic. Section \ref{sec:mehodology}, details the method applied in this study. Section \ref{sec:results}, presents the findings from each stage of the experiment. Section \ref{sec:discussion}, provides a general analysis and discussion of the results from all experiment stages. Limitations and future work directions are presented in Section \ref{sec:limitationsAndFutureWork}; and finally, the conclusion is in Section \ref{sec:conclusion}.

\vspace{-5pt}
\section{Background}\label{sec:background}
\subsection{Code Readability, Comprehensibility and SAT}
The topic of reading and understanding code has been studied for over 45 years \cite{wyrich_40_2024}. 
This activity can occupy 30 to 70\% of a developer's time \cite{minelli_i_2015}. 
According to \textcite{boehm_quantitative_1976}, code readability is a necessary characteristic for its comprehension. Codes measured as having low readability are identified as having low comprehensibility \cite{dantas_readability_2021}.

Code readability and comprehensibility are intertwined concepts as readability metrics are known to affect the code comprehension. Readability is a complex characteristic to measure due to its composition of various attributes \cite{buse_learning_2010, scalabrino_comprehensive_2018}. In their systematic mapping, \textcite{wyrich_40_2024} identified the most mentioned code attributes, as popular themes in mapped studies related to code comprehension. The most cited code attribute was identifier names (variable names, method names, etc.), mentioned in 16 studies. The second most cited attribute was code comments, present in 10 studies. These two attributes are the focus of our investigation.

Code readability is a desired metric for developers and their managers, for optimizing the decision making. This metric can inform decisions on which code should be prioritized for testing, refactoring, or detailed inspection during a code review performed by a human developer \cite{buse_information_2012}.

Static analysis software measures some readability characteristics of code like adherence to the conventions of the programming language in use. However, some other characteristics are measured with limitations. Documentation is measured by its proportion relative to the code, given as a percentage. The ability of the documentation to facilitate code comprehension is not evaluated. Even naming conventions for identifiers are limited to length (very short names) and dictionary use, without evaluating the meaning of the identifiers' names and its relevance to the program's context. These limitations are addressed only during the code review stage and are not automated \cite{gunawardena_concerns_2023}. An investigation about how LLMs' react to such attribute changes could unveil a path towards such automation.

\subsection{Large Language Model}
Large Language Models are Artificial Intelligence models that can have hundreds of billions of parameters, trained using a large volume of data. Commercial and open-source \ac{LLM} models such as ChatGPT, Gemini, Claude, and Llama use the \textit{Transformer} architecture \cite{vaswani_attention_2017}.

\ac{LLM}s have a significant capacity to learn from a few examples, in an approach called \ac{FSL}, where information about the task to be performed is provided in the command prompt. These models have shown surprising results even in \ac{ZSL} approaches, where no prior examples are provided; only a contextualization and description of the task to be performed are given in the prompt \cite{wang_large_2023}.

Between 2017 and 2024, more than 70 different \ac{LLM}s were used in research, being applied to various Software Engineering tasks. The most commonly used prompt engineering techniques are \ac{FSL} and \ac{ZSL} \cite{hou_large_2024}.

These models' learning abilities from a few examples are due to various factors, including their number of parameters, computational power, and the dataset used in their training. Historically, natural language processing (NLP) software was developed for specific tasks, which determined its modeling and training. These \ac{FSL} abilities, called as \textit{`emergent abilities'} by \textcite{wei_emergent_2022}, have allowed the software to be used for tasks beyond its original design. Furthermore, new \ac{LLM} models have become capable of surpassing results considered the state of the art for specialized models \cite{wei_emergent_2022}.

\vspace{-5pt}
\section{Related work}\label{sec:relatedWork}
This work lies at the intersection of \ac{LLM}s' and code readability and comprehensibility. We conducted a comprehensive review of literature exploring these intersections. Firstly, we examined automated approaches for measuring code readability and comprehensibility. Secondly, we investigated how \ac{LLM}s are utilized to enhance code readability and comprehensibility. To conduct this review, we used the results of the systematic mappings from \textcite{bexell_software_2020}, \textcite{wyrich_40_2024}, and \textcite{hou_large_2024}.

\textcite{bexell_software_2020} mapped 76 studies related to code readability, two of which specifically address the automation of its measurement.

\textcite{wyrich_40_2024} mapped 95 studies on code comprehensibility, two of which study its automation, with one being the same found in \textcite{bexell_software_2020}'s mapping. Other studies have also conducted some form of automation related to source code quality \cite{buse_learning_2010, dorn_general_2012, scalabrino_automatically_2019}.

\textcite{hou_large_2024} mapped 395 studies on the application of \ac{LLM} in various software engineering tasks, eight of which were classified as "Code understanding" research. Three studies investigate or propose improvements in \ac{LLM}s to enhance their performance in code-related activities, but without aiming to use \ac{LLM}s for code comprehension \cite{zhao_understanding_2023, wong_natural_2023, wang_codet5_2023}. A fourth study presents a dataset for training \ac{LLM}s, demonstrating its effectiveness in improving models' capabilities in code generation and comprehension tasks \cite{manh_vault_2023}.

We classified the mapped studies, focusing on automated approaches. The studies are grouped into three main categories: (1) Machine Learning Based Approaches and (2) Deep Learning Based Approaches.

\textbf{Machine Learning Based Approaches:} This group focuses on using quantifiable metrics extracted from the source code, documentation, and developer profiles to predict or explain code comprehensibility. These studies often employ statistical analysis and machine learning techniques to identify correlations and build predictive models. \textcite{buse_metric_2008} introduced a metric for software readability using the average score provided by 120 human annotator to train their binary classifier. \textcite{dorn_general_2012} also identified promising results, when compared with previous metric, on automating the measurement of code readability using also linguistic attributes (e.g., validating identifier names with English dictionaries). Further work also explored lexical artifacts \cite{scalabrino_improving_2016}. \textcite{scalabrino_automatically_2017} demonstrated that no single code attribute can be used to explain code comprehensibility in isolation. \textcite{trockman_automatically_2018} developed a binary logistic regression model, reanalyzing the data from \textcite{scalabrino_automatically_2017}, considering source code properties, documentation, and developer profile data (e.g., experience). This study highlighted the importance of combined metrics and revealed that code understandability would be measurable in future. \textcite{scalabrino_automatically_2019} investigated 121 metrics related to code, documentation, and developers. They found weak correlations between individual metrics and perceived comprehensibility. Although combining metrics showed some discriminatory power, the results were insufficient for practical use \cite{scalabrino_comprehensive_2018}. Additionally, those studies lack the evaluation of semantic aspects, such as meaning of identifier names and the meaning of the text in a comment, in relation to the code.

\textbf{Deep Learning Based Approaches:} This group utilizes Natural Language Processing (NLP) and other deep learning techniques to analyze code and related text (e.g., comments, documentation) to understand code semantics and improve comprehensibility. \textcite{kanade_learning_2020} presented CuBERT (Code Understanding BERT), achieving superior results in various code-related tasks, with one related in correlating comments with code (docstring task). \textcite{shen_benchmarking_2022} benchmarked pre-trained models (including CuBERT and CodeBERT) for identifying syntactic structures in code. They found that these models still lacked comprehensive code syntax understanding, achieving results inferior to simpler reference models. This highlights the challenges of applying NLP models to capture the nuances of code syntax.

\textcite{mi_inception_2018} introduced the use of \ac{CNN} trained on a matrix representation of code, generated by its characters ASCII values, inspired by image recognition approaches. Their model was compared to previous ML based models and achieved a slightly improved accuracy compared to the best one, from \textcite{scalabrino_improving_2016}. Latter \textcite{mi_improving_2018} presented a \ac{CNN} trained on code representation based on three levels of information: character-level (ASCI value), token-level (key word) and node level (\ac{AST}). Their model demonstrated improved compared to previous works on readability classification, including their previous model.

Seeking for improvements \textcite{mi_towards_2022} presented a novel approach using \ac{CNN} to extract structural features from the ASCII matrix representation of code. Also a combination of \ac{CNN} and BERT model to generate token representation and extract semantic features from code. Finally they have used a \ac{CNN} to extract visual features from RGB matrix generated from source code screen shots. Their new approach obtained the best performance compared to all previous, leading to the hypothesis that different deep learning technics combined can extract unmapped readability features. They also demonstrated that visual features contributed the least to the readability assessment, while the structural features being the most relevant.

\textbf{LLM-Based Approaches:} This group explores the use of LLMs for code comprehension and readability. \textcite{simoes_evaluating_2024} investigated the use of LLMs for source code quality assessment, including readability, by comparing the results with SonarQube analysis. Their study shows the LLM potential for evaluating code quality, but indicates the need for further studies. \textcite{hu_how_2024} found that LLMs are highly sensitive to semantic perturbations in code. However, their work did not seek to develop a metric based on this property.  \textcite{vitale_personalized_2025} evaluated the use of LLM to generate a personalized readability assessment, by using previous works' datasets \cite{buse_learning_2010, dorn_general_2012, scalabrino_comprehensive_2018} and adopting three labels "unreadable", "readable", "neutral". They compared their LLM approach with Scalabrino's Classifier, using precision, recall and f-1 score metrics. Their results demonstrated that the use of LLMs for personalized readability are less effective than state-of-art generalist model.

Studies related to code readability and comprehensibility have seen a progression from classifiers using machine learning algorithms \cite{buse_learning_2010} to NLP models \cite{kanade_learning_2020}, and more recently Deep Learning \cite{mi_enhanced_2022}. Current research focuses on evaluating the use of LLMs for evaluate code quality \cite{simoes_evaluating_2024} and readability \cite{vitale_personalized_2025}. 

Our work falls on LLM-Based Approach. The findings in this work can be used to direct further investigation towards the development of a code readability metric, or even a more broad novel code quality metric.

\section{Methodology}\label{sec:mehodology}

\newcommand{\itemA}{Subjects - Algorithms for analysis}%
\newcommand{\itemB}{Analysis of the original code with comparison group}%
\newcommand{\itemC}{Analysis of the original code with the LLMs}%
\newcommand{\itemD}{Interventions in the algorithms}%
\newcommand{\itemE}{Analysis after removing comments}%
\newcommand{\itemF}{Analysis after using confusing names}%
\newcommand{\itemG}{Analysis after code smell fix}%

In this work, we study the response of a set of objects (nine \ac{LLM}s) to the same set of interventions (code interventions) into the same set of subjects (12 Java Classes). The tested hypothesis consider that \ac{LLM}s will change its evaluation of source code accordingly to the interventions. There is no random selection of interventions, objects, or subjects, thus we are conducting a quasi-experiment \cite{shadish_experimental_2001}. This is a type of experiment often used in software engineering research \cite{wohlin_experimentation_2012}.

Therefore, we conducted a \textit{"blocked subject-object"} study, as we have nine study objects (\ac{LLM}s) and 12 subjects (12 classes) that will undergo changes (intervention) in their characteristics commonly related to readability and source code quality, which will be measured by the \ac{LLM}s \cite{wohlin_experimentation_2012}. 

\textcite{fakhoury_improving_2019} identified that tools used to detect style problems are able to capture readability characteristics of source code. In their study, those type of tools demonstrated better sensibility than models proposed by \textcite{scalabrino_improving_2016} and \textcite{dorn_general_2012}. The use of a comparison group is necessary to identify random changes on \ac{LLM}s' metric that are not caused by source code changes, but inherent LLM's variability.

We included two approaches as comparison groups \cite{shadish_experimental_2001}. One approach is the static analysis software SonarQube and some of its metrics. We chose this tool because it is broadly used in industry and some of its metrics evaluate code style problems. The other comparison reference is the \textcite{scalabrino_comprehensive_2018} model, which combine several source code features and had its accuracy extensively evaluated in previous works\cite{fakhoury_improving_2019, scalabrino_automatically_2019, sergeyuk_reassessing_2024}. Their reproducibility toolkit\footnote{https://dibt.unimol.it/report/readability/} has a working software, able to automatically extract features and perform code readability classification, without human intervention. This characteristic ensures that we do not introduce interpretation errors when building our own version of their model.

Yet, according to \textcite{shadish_experimental_2001}, the use of control groups are of minimal advantage unless they are also accompanied by pretest measurements taken on the same outcome variable as the posttest. So we executed a pretest, i.e, all source code was evaluated by SonarQube, Scalabrino Classifier and LLMs' before any code intervention, mitigating selection bias for both, source code and LLMs', creating a baseline metric.

The submission of the classes and their interventions to the \ac{LLM}s' analyses was not randomized. The interventions conducted on code were not random since they were guided by the measurements of the control measurement tool. Also, it was not necessary any randomized LLM analysis, since the LLM's had no context about previous code intervention, being unable to retain this information and compare the different states of code, which could affect its evaluation.

The LLM's were instructed to output its evaluation similarly to a binary classification, i.e. emit a 0 to 100 score. This approach ensures a direct comparison with the Scalabrino Classifier. Details about this approach are in Section \ref{sec:methLLM}.

\subsection{Comparison Group - SonarQube Metrics}

SonarQube has a set of rules known as Code Smells.
According to its documentation, Code Smells are issues related to maintainability.

Currently, SonarQube is replacing this definition with a more granular one called Clean Code,
which includes the following attributes: consistency, intentionality, adaptability, and responsibility. The first two attributes relate to code readability characteristics.

According to SonarQube's documentation "consistent code is formatted, conventional, and identifiable" and "intentional code is clear, logical, complete, and efficient".

SonarQube also measures the percentage of documentation lines relative to the code lines of a class, as well as the total number of lines in the class. 

In this experiment, we manipulate code attributes that affect metrics related to its readability, which can be captured by SonarQube's Clean Code metrics and documentation-related metrics directly measured by the percentage of documentation. 

To use SonarQube as our comparison group for metrics, we utilize Clean Code metrics and their attributes: Consistency (C), Intentionality (I), Adaptability (A) and Responsibility (R) representing the number of issues detected into the analyzed source code. Higher values for (C) and (I) are supposed to reduce source code readability. The percentage of comments or documentation (D), and the total number of lines (L) in each manipulated class as the SonarQube's documentation metric is a ration of the total lines of code. We refer to all Clean Code violations as Code Smells.

\subsection{Comparison Group - Scalabrino's Classifier}

As an approach evaluated against several datasets and also compared against code style tools, with an outstanding performance, we will take Scalabrino's classifier as our ground truth metric. It was tested against 600 code snippets, evaluated by 5k+ people. 

The classifier provided by \textcite{scalabrino_comprehensive_2018} works as a command line tool that receives either a code snippet with a Java method, an entire class file or a folder with any of these types. Upon its execution it produces a score between 0 (surely unreadable) and 1 (surely readable), rating the code readability.

The provided tool uses a \ac{OSS} library to parse an \ac{AST}. As its last update was in 2021, in order to make the tool work with current Java syntax, we updated the \ac{AST} parser library. The model combines structural and textual features. Comments Readability, Textual Coherence and Number of Concepts are the three most important textual features for this model. That makes it as the right fit to evaluate the type of code interventions we have performed, further explained in section \ref{sec:\itemD}.

The scores obtained with this classifier will be named as \ac{SC}. The classifier's output is a number with 16 digits precision, which we convert to the same LLM scale (0 to 100), with one digit precision.

\subsection{Objects - Selection of LLMs}\label{sec:methLLM}

We studied nine objects, which are the \ac{LLM}s: ChatGPT (4o and 4o-mini), Gemini (2.0 Pro and Flash), Llama (3.1 405B and 3.1 8B), Claude (3.7 Sonnet and 3.5 Haiku) and \deepSeek. Each of the first four listed \ac{LLM}s has both a robust and a fast version, except \deepSeek.

The objective is to evaluate the capability and sensitivity of the \ac{LLM}s in detecting changes in code attributes that affect its readability, compared to the detection made by the control tool, SonarQube.

There are various benchmarks for \ac{LLM}s, covering a wide range of applications and tasks, from general to more specific purposes \cite{chang_survey_2024}. This study does not aim to create or conduct a benchmark but rather to evaluate the performance of these models in assessing code readability.

However, given the constant evolution of commercial and open-source models, selecting models for research use becomes challenging. To select the best and most current models, we referred to the ranking on the Artificial Analysis website\footnote{Artificial Analysis https://artificialanalysis.ai/}, which classifies models offered through rest API, based on three main comparisons: Intelligence, Speed, and Price.

In this work, we did not analyze costs, so we selected the top four models in quality and the top four models in speed. If a model appeared in more than one criterion, we selected the next model in the ranking where the duplicated model had the least affinity.

At the time of planning this work, the top four models in quality and speed are the ones listed in Table \ref{tab:melhoresModelos}. We did not considered the models classified as reasoning model. The \gptMini\ model was also ranked second by quality, but since its purpose is speed and it is present in the respective ranking, we removed it from the intelligence list. Similarly \geminiPro\ ranked third on speed, but we considered it on intelligence rank only.

During execution of this research (between 10/2024 and 03/2025), some models received updates and the \deepSeek model entered the rank. We decided to execute the update and to add this new model.

\begin{table}[ht]
    \centering
    \scriptsize
    \captionof{table}{Selected models}
    \label{tab:melhoresModelos}
    \begin{tabular}{lrrl}
        \toprule
        \textbf{Model} & \textbf{Score} & \textbf{Release Date} & \textbf{Criteria} \\
        \midrule
        \geminiPro & 49 & 05/02/2025 & intelligence \\
        \claudeSonnet & 48 & 19/02/2025 & intelligence \\
        \deepSeek & 46 & 26/12/2024 & intelligence \\
        \gptO & 41 & 13/05/2024 & intelligence \\
        \llamaQuatrocentos & 40 & 23/07/2024 & intelligence \\
        \geminiFlash & 265 & 05/02/2025 & speed \\
        \llamaOito & 185 & 23/07/2024 & speed \\
        \gptMini & 79 & 18/07/2024 & speed \\
        \claudeHaiku & 66 & 13/03/2024 & speed \\
        \bottomrule
    \end{tabular}
    \vspace{-5pt}
\end{table}

All models were used through REST APIs from the following sites: OpenAI\footnote{OpenAI https://openai.com/}, Google\footnote{Google AI Studio https://ai.google.dev/aistudio}, Anthropic\footnote{Anthropic https://www.anthropic.com/}, DeepInfra\footnote{DeepInfra https://deepinfra.com/} and DeepSeek\footnote{DeepSeek https://deepseek.com}.

These models are able to learn from a few examples. This capability is due to various factors, including their number of parameters, computational power, and the dataset used in their training. These \ac{FSL} abilities, referred to as \textit{emergent abilities} by \textcite{wei_emergent_2022}, have enabled the software to perform tasks beyond its original design. Furthermore, new \ac{LLM}s have become capable of surpassing results considered the state-of-the-art for specialized models \cite{wei_emergent_2022}. In this work, we adopt a \ac{ZSL} approach.

To create the prompt that guides the \ac{LLM} on how to respond to the assigned task, we used three patterns mapped by \textcite{white_prompt_2023} (\textit{Persona, Template, and Reflection}). We instructed the \ac{LLM} to assume a persona capable of performing the analysis with a certain knowledge. For this purpose, we used the passage: \textit{"The assistant is a seasoned senior software engineer, with deep Java Language expertise, doing source code evaluation as part of a due diligence process, these source codes are presented in the form of a Java Class File. Your task is to emit a score from 0 to 100 based on the readability level of the source code presented."}. 

The \ac{LLM} was requested to present an explanation for the grade, by passing guidelines on its format \textit{"- The ”explanation” attribute must not surpass 450 characters and MUST NOT contain special characters or new lines."}. To enable the \ac{LLM}'s responses to be processed via software, we requested it to follow a template for the response: \textit{"Your answers MUST be presented ONLY in the following json format: {”score”:”NN\%”, reasoning:”your explanation about the score”}"}. This explanation was used for debug purposes to iterate between prompts improvements.

In all executions, the prompt remained unchanged. Each class was analyzed 10 times, totaling 120 analyses and consuming about 1.2 million tokens per model on each scenario. The total number of analysis for this research was 4.320 or around 43.2 million tokens. The results were submitted to an analysis of the variability of the responses presented by each \ac{LLM}. The output files of all executions, as well as the Python code used to perform these analyses, are preserved in a GitHub repository
\makeatletter
\if@ACM@anonymous
    \textit{(reference hidden for anonymization purposes)}.
\else
    \cite{LLMSonarQuarkusAnalysis}.
\fi
\makeatother 

\subsection{\itemA}\label{\itemA}
The most commonly used language in studies related to code readability and comprehensibility is Java, with 47\% of published articles. Over 40 years of research, more than 88\% of studies have used only one programming language in their investigations. The most common approaches for obtaining code snippets for readability studies were: (1) code produced by researchers, (2) code from open-source projects, and (3) code used in previous works \cite{wyrich_40_2024}. Also, \textcite{scalabrino_comprehensive_2018} model has proven results only against Java code, as well as its toolkit is able to extract features only from Java \ac{AST}.

For such reason, we decided to use Java language software in this research and selected the open-source Open JDK project\footnote{Open JDK https://github.com/openjdk/jdk}, which refers to the fundamental libraries of the Java SDK. We selected two classes with rich API documentation, with the intention to clearly detect the impact of a rich documentation removal from the sour code. The two selected classes were \textit{java.util.DoubleSummaryStatistics} and \textit{java.time.Month}. 

Additionally, a set of demonstration classes from the JDK was selected. This set of classes was chosen for their less rich documentation since they were written to demonstrate the language's use. The list of selected classes is as follows: root package ../demo/share/jfc/, sub packages SampleTree, classes DynamicTreeNode, SampleData, SampleTree, SampleTreeCellRenderer, SampleTreeModel, sub package Notepad, classes ElementTreePanel, Notepad, sub package Stylepad, classes HelloWorld, Stylepad and Wonderland.


The total number of original selected classes is 12, totaling 4,051 lines. Each class was used in 4 different intervention states, totaling 48 classes with 14.919 lines. The number of classes and code was kept low to allow for manual interventions in the code. The main studies in this area have used the following datasets: \textcite{buse_learning_2010} dataset with 100 snippets totaling 769 lines of code, \textcite{dorn_general_2012} has 120 snippets totaling 3,617 lines and \textcite{scalabrino_comprehensive_2018} has 200 snippets totaling 5.140 lines. \textcite{sergeyuk_reassessing_2024} has 120 snippets totaling 5,533 lines.


\subsection{\itemD}\label{sec:\itemD}

To explore how source code attributes affect the scores assigned by the \ac{LLM}s, we perform interventions on the code, followed by re-analyses by SonarQube and the \ac{LLM}s. After each intervention, we present the results from the \ac{LLM}s. \textbf{Identifier naming} and \textbf{code comments} are the two most studied source code attributes related to code readability and comprehensibility \cite{wyrich_40_2024}. The interventions will target these two attributes. We did not test if the changes cause an improvement or deterioration to readability, other than compare the results with the evaluation of comparison group.

\textit{Intervention on code comments:} In the second analysis, we identify the impact of comments on SonarQube and \ac{LLM} evaluations by removing the comments and re-running the analysis. All comments were removed except source code that was commented out.

\textit{Intervention on variable names:} In the third analysis, we alter the names of variables and methods. Good variable and method names facilitate code comprehension. This is a widely accepted axiom by industry \cite{fowler_refactoring_1999} and extensively investigated showing that consistent use of descriptive names improve code readability \cite{wyrich_40_2024, al_madi_novice_2021, schankin_descriptive_2018, scalabrino_improving_2016, buse_learning_2010}.

In order to emulate such an extreme intervention as the removal of all comments, we modify all variables and methods to semantically confusing values, such as names of vegetables and spices (potato, tomato, parsley, rosemary etc). 

We expect SonarQube not to show changes in its evaluation, but the \ac{LLM}s should be able to identify the confusion caused in the code, impacting the score. Class names are preserved to allow comparison of results, as well as overridden methods.

\textit{Intervention on Code Smells:} In the fourth analysis, we fix all fixable Code Smells, especially those of Clean Code Attributes C and I, that are strongly related to readability.

We expect these changes to positively impact SonarQube metrics as well as the \ac{LLM}s' evaluations. Changes are performed only on methods and classes containing a Code Smell. Thus, classes $C_1$, $C_7$ and $C_{10}$, which do not have Code Smells, do not undergo intervention at this stage, as can be seen on Table \ref{tab:tabelaEstatisticasSonar}, detailed on Section \ref{\itemB}.

\textit{Code versions:} The interventions generate four code versions for the 12 classes (with the identifier we use to refer to each scenario in parentheses): 
\begin{itemize}
    \item [] (OC) Original code state;
    \item [] (I1) without comments;
    \item [] (I2) with confusing names for identifiers;
    \item [] (I3) with fixed Code Smells. 
\end{itemize}

In Section \ref{sec:discussion}, we compare the results of all analyses.

Every code analysis involved statistical evaluation of the LLM performance.  For each scenario ($S \in \{OC, I1, I2, I3\}$) and Java class ($C_i$, where $i \in \{0, 1, \ldots, 12\}$), we calculated the mean and standard deviation of the scores from 10 analysis samples for each LLM.

To compare LLM performance across scenarios for each class, we computed the precision, recall and f-1 score between LLM scores and the Scalabrino Classifier (SCO), taken as our ground truth, as previous work \cite{vitale_personalized_2025}. To stablish this ground truth, we have applied the same interval found by \textcite{piantadosi_how_2020} for the SCO. In their work they have found that the readability prediction tool is unreliable when its output score is within the range of [0.416, 0.600]. Based on this, we established thresholds for classification: a score below 0.416 classifies the code as 'unreadable,' and a score above 0.600 classifies it as 'readable.' Scores within this interval are considered 'neutral'. We applied the same thresholds to all LLM's scores.

\subsection{Thematic Coding of LLM Reasoning}\label{sec:itemE}

To systematically analyze the LLMs reasoning for the scores they assigned, we employed an inductive thematic analysis methodology. This approach, as outlined by \textcite{guest_applied_2012}, is designed to identify and examine themes from textual data in a transparent, credible, and systematic manner. The primary objective of this analysis was to understand the qualitative aspects and the criteria the LLMs used to evaluate code readability across the different intervention scenarios. The process was executed as follows:

\textbf{Data Preparation and Segmentation:} Comments from all LLMs for a given scenario have been consolidated into a single file, containing one comment per line.

\textbf{Inductive Theme Identification:} We did not start with a preconceived set of themes. Instead, two researchers independently read a subset of the LLM reasoning texts. The goal was to become familiar with the data and to notice recurring concepts, ideas, and justifications---the emergent themes.

\textbf{Codebook Development:} After the initial pass, a preliminary list of themes was discussed and refined. This led to the creation of a formal codebook. For each code, we developed \textbf{A descriptive label:} A short mnemonic for easy reference (e.g., \texttt{Good Readability}) and \textbf{Inclusion/Exclusion Criteria:} A set of keywords or short expressions to be used for regular expression matching.
    
\textbf{Coding and Inter-Coder Agreement:} For each thematic code in the codebook, a corresponding set of regular expression patterns was developed to identify its presence within the textual data. To avoid false positives in cases of ambiguous explanations, whenever a negative code (e.g., \texttt{Poor Readability}) was detected, the corresponding positive code (e.g., \texttt{Good Readability}) was not accounted for in the same comment. 

\textbf{Data Reduction and Analysis:} The final step was to analyze the resulting structured data. This was achieved by quantifying the frequencies of each thematic code for each of the four scenarios (OC, I1, I2, and I3).

This structured and inductive approach ensured that the analysis of the LLMs' reasoning was not merely a subjective interpretation but a methodologically sound process, grounded in the data itself and aligned with established practices in qualitative data analysis.

\section{Results}\label{sec:results}

\subsection{\itemB}\label{\itemB}
The metrics shown in Table \ref{tab:tabelaEstatisticasSonar} were collected to be used as the basis for comparison with the \ac{LLM}s' results. A total of nine Code Smells were identified by SonarQube. All classes were rated A for maintainability, indicating a high presumed quality level. All classes are referenced in this section by its alias, as listed in table \ref{tab:tabelaEstatisticasSonar}.

\begin{table}[ht]
\centering
\scriptsize
\caption{SonarQube attributes values}
\label{tab:tabelaEstatisticasSonar}
\begin{tabular}{lrrrrrrrr}
    \toprule
    \textbf{Alias (class name)} & \multicolumn{4}{c}{\textbf{Clean Code}} & \multicolumn{1}{c}{\textbf{D}} & \multicolumn{1}{c}{\textbf{L}}& \multicolumn{1}{c}{\textbf{SC}}\\
        & \textbf{C} & \textbf{I} & \textbf{A}&\textbf{R} & & \\
    \midrule
    $C_1$ (DoubleSummaryStatistics)   &-&-&-&-&62.9\%&294&90.7\\
    $C_2$ (Month)                     &2&1&-&-&63.4\%&526&91.0\\
    $C_3$ (DynamicTreeNode)           &-&2&-&-&32.1\%&155&93.3\\
    $C_4$ (ElementTreePanel)          &-&3&2&-&23.0\%&579&78.9\\
    $C_5$ (HelloWorld)                &-&3&2&-&11.0\%&177&64.4\\
    $C_6$ (Notepad)                   &9&20&5&-&10.2\%&824&77.4\\
    $C_7$ (SampleData)                &-&-&-&-&25.5\%&75&93.3\\
    $C_8$ (SampleTree)                &-&5&5&-&27.4\%&596&74.8\\
    $C_9$ (SampleTreeCellRenderer)    &-&1&1&-&16.5\%&131&80.2\\
    $C_{10}$ (SampleTreeModel)          &-&-&-&-&38.2\%&48&89.1\\
    $C_{11}$ (Stylepad)                 &-&15&7&-&7.9\%&378&74.0\\
    $C_{12}$ (Wonderland)               &-&1&6&-&5.7\%&268&68.5\\
    \bottomrule
\end{tabular}
\vspace{-5pt}
\end{table}

Only three of the 12 classes do not have any Code Smell. Most of the Code Smells are of type C and I, strongly related to source code readability according to SonarQube documentation. The \textit{$C_1$} and \textit{$C_2$} classes have more than 60\% of comments, and all classes have some level of comments. The complexity indicators show that some of the classes have a low level of complexity while others have higher levels, yet still considered low.

\subsection{\itemC}\label{\itemC}

Next, we list the models that presented variation on score attribution, with their respective standard deviation (SD) and coefficient of variation (CV), using notation $C_i$ SD (CV): 

\begin{description}[font=\small]
    \item[\gptO]- $C_7$ 1.6 (1,67\%)
    \item[\gptMini]- $C_4$ 4.2 (5,08\%)
    \item[\llamaQuatrocentos]- $C_4$ 2.4 (2,89\%)
    \item[\geminiPro]- $C_4$ 4.8 (6.62\%), $C_6$ 1.6 (2.24\%), $C_9$ 2.6 (3.01\%) and $C_{11}$ 2.6 (3.59\%).
\end{description}
Half of the models did not show any variation in the scores assigned.

The $C_1$ and $C_2$ classes, which are rich in documentation, received the highest evaluations from the \ac{LLM}s. 

As shown in Table \ref{tab:agreementOriginal}, all nine evaluated LLMs demonstrated 100\% agreement with the Scalabrino Classifier (SC) for the original code scenario. In this scenario, all classes were classified as readable by SCO and by all LLM.

\newcommand{\ccr}{\cellcolor{red!15}}
\newcommand{\ccg}{\cellcolor{green!15}}
\newcommand{\ccy}{\cellcolor{yellow!20}}
\newcommand{\ccc}{\cellcolor{gray!25}}
\newcommand{\ccb}{\cellcolor{blue!15}}

\begin{table}[htbp]
\vspace{-5pt}
    \centering
    \scriptsize
    \caption{Agreement with Scalabrino Classifier - Original Code}
    \begin{tabular}{lrrrrrrrrr}
        \toprule
        & \multicolumn{2}{c}{\textbf{ChatGPT}} & \multicolumn{2}{c}{\textbf{Gemini}} & \multicolumn{2}{c}{\textbf{Llama}}& \multicolumn{2}{c}{\textbf{Claude}}&\textbf{Deep}\\
        \cmidrule{2-9}
        \textbf{Metric} & 4o & 4o & 2.0 & 2.0 & 3.1 & 3.1 & 3.7 & 3.5 & \textbf{Seek} \\
                  &   & mini & Pro & Flash & 405B & 8B & Sonnet & Haiku & V3 \\
        \midrule
        Precision & 100\% & 100\% & 100\% & 100\% & 100\% & 100\% & 100\% & 100\% & 100\% \\
        Recall & 100\% & 100\% & 100\% & 100\% & 100\% & 100\% & 100\% & 100\% & 100\% \\
        F1-Score & 100\% & 100\% & 100\% & 100\% & 100\% & 100\% & 100\% & 100\% & 100\% \\
        \bottomrule
    \end{tabular}
    \label{tab:agreementOriginal}
    \vspace{-5pt}
\end{table}

\subsection{\itemE}\label{\itemE}
In the second analysis, Sonar's rating remained A, but the number of Code Smells increased due to the repercussion of removing comments, as classes with public methods lacking JavaDoc, violating a quality rule\footnote{SonarQube Rule: https://rules.sonarsource.com/java/RSPEC-1176/}, or empty blocks of code, that should at least be documented\footnote{SonarQube Rule: https://rules.sonarsource.com/java/RSPEC-108/}. 

In Table \ref{tab:tabelaEstatisticasSonarComments}, we marked the measures reductions in red and increments in green. The only Code Smell reduction was due to the removal of a comment with a FIXME string that was generating a Code Smell by SonarQube, indicating the need to developer to take an action\footnote{SonarQube Rule: https://rules.sonarsource.com/java/RSPEC-1134/}. As expected, the total lines of code (L) and documentation ratio (D) decreased, except for the $C_{5}, C_{8}, C_{11}$ classes, which contain commented-out source code that was kept after the intervention. The \ac{SC} also decreased the score, but not for class $C_{5}$.

\begin{table}[ht]
\centering
\scriptsize
\caption{SonarQube values after comments removal}
\label{tab:tabelaEstatisticasSonarComments}
\begin{tabular}{lrrrrrrrr}
    \toprule
    \textbf{Class} & \multicolumn{4}{c}{\textbf{Clean Code}} & \multicolumn{1}{c}{\textbf{D}} & \multicolumn{1}{c}{\textbf{L}}& \multicolumn{1}{c}{\textbf{SC}} \\
        & \textbf{C} & \textbf{I} & \textbf{A}&\textbf{R} & & \\
    \midrule
    $C_1$ &-&\ccg1&-&-&\ccr0.0\%&\ccr99&\ccr80.0\\
    $C_2$ &2&1&-&-&\ccr0.0\%&\ccr177&\ccr77.1\\
    $C_3$ &-&2&-&-&\ccr0.0\%&\ccr103&\ccr82.1\\
    $C_4$ &-&3&2&-&\ccr0.0\%&\ccr424&\ccr59.7\\
    $C_5$ &-&\ccg4&2&-&\ccr1.4\%&\ccr166&\ccg71.5\\
    $C_6$ &9&\ccr19&5&-&\ccr0.0\%&\ccr699&\ccr71.1\\
    $C_7$ &-&-&-&-&\ccr0.0\%&\ccr47&\ccr87.7\\
    $C_8$ &-&5&5&-&\ccr5.0\%&\ccr451&\ccr61.7\\
    $C_9$ &-&1&1&-&\ccr0.0\%&\ccr109&\ccr47.1\\
    $C_{10}$&-&-&-&-&\ccr0.0\%&\ccr28&\ccr69.4\\
    $C_{11}$&-&15&7&-&\ccr1.3\%&\ccr239&\ccr70.5\\
    $C_{12}$&-&\ccg2&6&-&\ccr0.0\%&\ccr69&\ccr61.9\\
    \bottomrule
\end{tabular}
\vspace{-5pt}
\end{table}

Below, we list models that presented variation on score attribution in this scenario, along with their respective SD and CV:

\begin{description}[font=\small]
    \item[\claudeSonnet]- $C_{10}$ 5.2 (6.54\%);
    \item[\claudeHaiku]- $C_1$ 2.6 (2.97\%);
    \item[\llamaOito]- $C_4$ 6.3 (8.11\%);
    \item[\llamaQuatrocentos]- $C_1$ 2.2 (2.58\%) and $C_7$ 2.6 (3.11\%);
    \item[\deepSeek]- $C_1$ 2.6 (2.85\%) and $C_7$ 2.1 (3.32\%);
    \item[\geminiPro]- $C_5$ 3.2 (4.16\%), $C_8$ 1.6 (2.28\%), $C_{10}$ 2.6 (3.01\%), $C_{11}$ 2.1 (2.97\%) and $C_{12}$ 1.3 (1.63\%).
\end{description}

Table \ref{tab:scoreMedioPorClasseELLMNoComments} lists the variation between the results of the first and second analyses. The new distribution of \ac{LLM} scores showed 38 score reductions and nine score increases. We marked the reductions in red, increases in green, changes with moderate statistical significance in yellow and with no statistical significance in gray.

\begin{table}[hbt]
    \centering
    \tiny
    \caption{Average score by class and LLM - Comments removed}
    \label{tab:scoreMedioPorClasseELLMNoComments}
    \begin{tabular}{lrrrrrrrrrr}
        \toprule
 & \multicolumn{2}{c}{\textbf{ChatGPT}} & \multicolumn{2}{c}{\textbf{Gemini}} & \multicolumn{2}{c}{\textbf{Llama}}& \multicolumn{2}{c}{\textbf{Claude}}&\textbf{DeepSeek}&\textbf{SC}\\
        \cmidrule{2-10}
        \textbf{Class} 
            & 4o        & 4o-mini   & 2.0-Pro   & 2.0-Flash & 3.1-405B  & 3.1-8B      & 3.7-Sonnet & 3-Haiku&V3&\\
        \midrule
        $C_1$&0.0\%     &0\%        &\ccr-5.3\% &\ccr-5.6\% &\ccr-9.8\% &\ccr-13.0\% &\ccr-5.6\%&\ccr-8.4\%&\ccc-2.6&\ccr-11.82\%\\
        $C_2$&0.0\%     &\ccr-10.5\%&0.0\%      &0.0\%      &0.0\%      &0.0\%      &\ccr-5.3\% &0.0\%      &   0.0&\ccr-15.24\%\\
        $C_3$&0.0\%     &\ccr-11.8\%&0.0\%      &\ccr-11.8\%&\ccr-15.3\%&\ccr-34.8\%&\ccr-11.8\%&\ccr-23.5\%&   0.0&\ccr-12.00\%\\
        $C_4$&0.0\%     &\ccy-9.6\% &\ccc-4.1\% &0.0\%      &\ccy-4.2\% &\ccr-8.2\% &\ccr-11.8\%&0.0\%      &   0.0&\ccr-24.33\%\\
        $C_5$&0.0\%     &0.0\%      &\ccc1.3\%  &0.0\%      &0.0\%      &\ccr-25.0\%&0.0\%      &\ccg13.3\% &   0.0&\ccg11.00\%\\
        $C_6$&0.0\%     &0.0\%      &\ccc-0.7\% &0.0\%      &\ccg16.7\% &0.0\%      &\ccr-11.8\%&0.0\%      &  0.0&\ccr-8.14\%\\
        $C_7$&\ccr-10.1\%&0.0\%     &0.0\%      &\ccr-5.3\% &\ccc-2.4\% &0.0\%      &\ccr-11.8\%&\ccr-11.8\%&\ccy-4.2&\ccr-6.00\%\\
        $C_8$&0.0\%       &\ccr-11.8\%&\ccc0.7\%&0.0\%      &\ccr-12.5\%&0.0\%    &\ccr-11.8\%&\ccr-11.8\%& 0.0&\ccr-17.44\%\\
        $C_9$&0.0\%     &0.0\%      &\ccc-2.9\% &0.0        &\ccr-5.9\% &0.0\%      &0.0\%      &\ccr-11.8\%&   0.0&\ccr-41.26\%\\
        $C_{10}$&0.0\%  &0.0\%      &\ccc2.9\%  &\ccr-5.6\% &0.0\%      &\ccr-5.9\% &\ccg5.3\%  &0.0\%      &   0.0&\ccr-22.01\%\\
        $C_{11}$&0.0\%  &0.0\%      &\ccc-1.4\% &0.0\%      &0.0\%      &\ccr-25.0\%&0.0\%      &0.0\%      &   0.0&\ccr-4.72\%\\
        $C_{12}$&0.0\%  &0.0\%      &\ccy3.2\%  &0.0\%      &\ccg14.3\%&0.0\%       &\ccg15.4\% &0.0\%      &   0.0&\ccr-9.66\%\\
        \bottomrule
    \end{tabular}
    \vspace{-5pt}
\end{table}

All models presented at least one score variation, \gptO\ showed only one reduction, while \deepSeek\ had two - one with no statistical significance and the other with moderate significance. \claudeSonnet\ presented variations for nine classes, all statistically significant. Seven out of nine changes pointed by \geminiPro\ had no statistical significance, making it the model with the most noise.

Table \ref{tab:agreementNoComments} shows a uniform recall of 83.3\% across all nine LLMs, as they have successfully identified most of the classes in accordance with SC.  The precision is lower than the recall. This suggests that the LLMs incorrectly classified a number of unreadable classes (according to the SC) as readable, resulting in a higher rate of false positives. This was caused by 2 of the 12 classes being on neutral range by Scalabrino Classifier, while all classes remained classified as readable by all LLM.

\begin{table}[htbp]
\vspace{-5pt}
    \centering
    \scriptsize
    \caption{Agreement with Scalabrino Classifier - Comments Removed}
    \begin{tabular}{lrrrrrrrrr}
        \toprule
        & \multicolumn{2}{c}{\textbf{ChatGPT}} & \multicolumn{2}{c}{\textbf{Gemini}} & \multicolumn{2}{c}{\textbf{Llama}}& \multicolumn{2}{c}{\textbf{Claude}}&\textbf{Deep}\\
        \cmidrule{2-9}
        \textbf{Metric} & 4o & 4o & 2.0 & 2.0 & 3.1 & 3.1 & 3.7 & 3.5 & \textbf{Seek} \\
                   &   & mini & Pro & Flash & 405B & 8B & Sonnet & Haiku & V3 \\
        \midrule
        Precision & 69.44\% & 69.44\% & 69.44\% & 69.44\% & 69.44\% & 69.44\% & 69.44\% & 69.44\% & 69.44\% \\
        Recall & 83.33\% & 83.33\% & 83.33\% & 83.33\% & 83.33\% & 83.33\% & 83.33\% & 83.33\% & 83.33\% \\
        F1-Score & 75.76\% & 75.76\% & 75.76\% & 75.76\% & 75.76\% & 75.76\% & 75.76\% & 75.76\% & 75.76\% \\
        \bottomrule
    \end{tabular}
    \label{tab:agreementNoComments}
    \vspace{-5pt}
\end{table}

\subsection{\itemF}\label{\itemF}

In the third analysis, Sonar's rating remained A and no attributes changed, making this scenario invisible to Sonar's analysis. The new distribution of \ac{LLM} scores showed a noticeable decline compared to previous versions. This indicates that the naming of variables and methods in the code impacts the perception of readability according to the \ac{LLM} evaluations.

All models changed its variability behavior upwards, \geminiFlash, \claudeSonnet\ and \deepSeek\ presented no variation. Below we list the variation on score attribution in this scenario for the other models, with their respective SD and CV:

\begin{description}[font=\small]
    \item[\claudeHaiku]- $C_2$ 5.2 (5.67\%);
    \item[\llamaQuatrocentos]- $C_5$ 1.0 (2.38\%);
    \item[\llamaOito]- $C_3$ 6.3 (8.11\%) and $C_7$ 9.7 (13.06\%);
    \item[\gptMini]- $C_1$ 4.2 (5.78\%), $C_6$ 3.2 (4.94\%) and $C_{11}$ 4.2 (5.78\%);
    \item[\gptO]- $C_1$ 2.6 (3.80\%), $C_3$ and $C_4$ 5.2 (6.54\%);
    \item[\geminiPro]- $C_2$ 2.1 (2.24\%), $C_4$ 4.2 (6.30\%), $C_5$ 4.2 (19.17\%), $C_6$ 3.2 (10.20\%) and $C_9$, $C_{11}$ 1.6 (2.61\%);
\end{description}


Table \ref{tab:scoreMedioPorClasseELLMBadNames} presents the scores obtained by class and LLM for this analysis. Almost all evaluations showed a score reduction highlighted in red. \gptO\ presented two reductions with moderate statistical significance, in yellow. \gptMini\, \llamaOito\ and \claudeHaiku\ presented one non statistical significant reduction each one and \geminiPro \ presented two, all of them in gray color.

\begin{table}[hbt]
    \centering
    \tiny
    \caption{Average Score by Class and LLM - Use of Confusing Names}
    \label{tab:scoreMedioPorClasseELLMBadNames}
    \begin{tabular}{lrrrrrrrrrr}
        \toprule
         & \multicolumn{2}{c}{\textbf{ChatGPT}} & \multicolumn{2}{c}{\textbf{Gemini}} & \multicolumn{2}{c}{\textbf{Llama}}& \multicolumn{2}{c}{\textbf{Claude}}&\textbf{DeepSeek}&\textbf{SC}\\
        \cmidrule{2-10}
        \textbf{Class} 
            & 4o            & 4o-mini   & 2.0-Pro   & 2.0-Flash     & 3.1-405B & 3.1-8B    & 3.7-Sonnet    & 3.5-Haiku   &  V3  &\\
        \midrule
        $C_1$   &\ccr-28.4\%&\ccr-14.1\%&\ccr-68.4\%&\ccr-38.9\%   &\ccr-57.9\%&\ccr-7.6\% &\ccr-77.8\%    &\ccr-10.5\%&\ccr-57.9&\ccr-0.69\% \\
        $C_2$   &\ccr-10.5\%&\ccr-10.5\%&\ccc-1.1\% &\ccr-5.6\%    &\ccr-15.8\%&\ccr-3.1\% &\ccr-31.6\%    &\ccc-4.2\% &\ccr-10.5&\ccr-1.72\% \\
        $C_3$   &\ccy-7.1\% &\ccr-11.8\%&\ccr-29.4\%&\ccr-47.1\%   &\ccr-29.4\%&\ccr-15.2\%&\ccr-58.8\%    &\ccr-11.8\%&\ccr-11.8&\ccr-2.34\% \\
        $C_4$   &\ccy-7.1\% &\ccr-9.6\% &\ccc-7.5\% &\ccr-13.3\%   &\ccr-28.1\%&0.0\%      &\ccr-52.9\%    &\ccr-11.8\%&\ccr-23.5&\ccg0.13\%  \\
        $C_5$   &\ccr-11.8\%&\ccr-13.3\%&\ccr-70.7\%&\ccr-60.0\%   &\ccr-42.0\%&\ccr-25.0\%&\ccr-53.8\%    &\ccr-13.3\%&\ccr-52.9&\ccr-13.12\% \\
        $C_6$   &\ccr-23.5\%&\ccr-14.7\%&\ccr-56.0\%&\ccr-46.7\%   &\ccr-66.7\%&0.0\%      &\ccr-64.7\%    &\ccr-23.5\%&\ccr-47.1&\ccg0.85\%  \\
        $C_7$   &\ccr-20.6\%&\ccr-17.6\%&\ccr-26.3\%&\ccr-78.9\%   &\ccr-29.4\%&\ccc-7.5\% &\ccr-58.8\%    &\ccr-23.5\%&\ccr-31.6&\ccr-1.56\% \\
        $C_8$   &0.0\%      &\ccr-11.8\%&0.0\%      &\ccr-13.3\%   &\ccr-25.0\%&0.0\%      &\ccr-58.8\%    &0.0\%      &\ccr-23.5&\ccr-2.34\% \\
        $C_9$   &0.0\%      &0.0\%      &\ccr-30.9\%&\ccr-11.8\%   &\ccr-29.4\%&\ccr-25.0\%&\ccr-60.0\%    &\ccr-23.5\%&\ccr-11.8&\ccg1.11\%  \\
        $C_{10}$&0.0\%      &0.0\%      &\ccr-17.6\%&\ccr-5.6\%    &0.0\%      &0.0\%      &\ccr-20.0\%    &\ccr-11.8\%&\ccr-11.8&\ccr-0.50\% \\
        $C_{11}$&\ccr-23.5\%&\ccc-2.7\% &\ccr-16.0\%&\ccr-46.7\%   &\ccr-33.3\%&\ccr-25.0\%&\ccr-60.0\%    &\ccr-23.5\%&\ccr-23.5&\ccr-1.25\% \\
        $C_{12}$&\ccr-11.8\%&\ccr-13.3\%&\ccr-60.0\%&\ccr-40.0\%   &\ccr-42.9\%&\ccr-29.4\%&\ccr-53.8\%    &\ccr-23.5\%&\ccr-52.4&\ccr-10.06\% \\
        \bottomrule
    \end{tabular}
    \vspace{-5pt}
\end{table}

\geminiFlash, \claudeSonnet\ and \deepSeek\ were the only ones reducing scores for every class. The \llamaOito\ model maintained the score for the largest number of classes, followed by \gptO\ and \gptMini.

Table \ref{tab:agreementBadNames} reveals the increased and inconsistent divergence between \ac{LLM}s readability scores with a reference model. Unlike previous scenarios, the agreement rates vary dramatically between models. They range from a high recall of 91.67\% for models like \gptO, \gptMini, \llamaOito, and \claudeHaiku, to an extremely low 16.67\% for \claudeSonnet. SCO marked two classes as neutral and kept all others as readable.

\geminiFlash\ and \claudeSonnet~reacted most strongly to the nonsensical names. \claudeSonnet\ exemplifies this with an extremely low recall of 16.67\%, correctly identifying only two readable classes and marking all others as unreadable. This shows a deep semantic analysis that diverges sharply from the SC's assessment, resulting in a very low F1-Score (28.21\%).

\geminiPro\ and \deepSeek\ were more conservative. Their high precision (91.67\%) shows that when they classified code as readable, they were very likely to be correct according to the SC. However, their lower recall (66.67\%) means they were stricter than the SCO and flagged more code as unreadable.

Similar to previous scenario SCO marked almost all classes as readable, but one as neutral.

\begin{table}[htbp]
\vspace{-5pt}
    \centering
    \scriptsize
    \caption{Agreement with Scalabrino Classifier - use of confusing names}
    \begin{tabular}{lrrrrrrrrr}
        \toprule
         & \multicolumn{2}{c}{\textbf{ChatGPT}} & \multicolumn{2}{c}{\textbf{Gemini}} & \multicolumn{2}{c}{\textbf{Llama}}& \multicolumn{2}{c}{\textbf{Claude}}&\textbf{Deep}\\
        \cmidrule{2-9}
        \textbf{Metric} & 4o & 4o & 2.0 & 2.0 & 3.1 & 3.1 & 3.7 & 3.5 & \textbf{Seek} \\
                   &   & mini & Pro & Flash & 405B & 8B & Sonnet & Haiku & V3 \\
        Precision & 84.03\% & 84.03\% & 91.67\% & 91.67\% & 91.67\% & 84.03\% & 91.67\% & 84.03\% & 91.67\% \\
        Recall & 91.67\% & 91.67\% & 66.67\% & 41.67\% & 58.33\% & 91.67\% & 16.67\% & 91.67\% & 66.67\% \\
        F1-Score & 87.68\% & 87.68\% & 77.19\% & 57.29\% & 71.30\% & 87.68\% & 28.21\% & 87.68\% & 77.19\% \\
        \bottomrule
    \end{tabular}
    \label{tab:agreementBadNames}
    \vspace{-5pt}
\end{table}

\subsection{\itemG}\label{\itemG}
In the fourth analysis, SonarQube indicated a strong reduction in Code Smells, along with changes in comments and lines of code. In Table \ref{tab:tabelaEstatisticasSonarRefactoring} we marked the measures reductions in red and increments in green. Four classes suffer a score reduction by our reference classifier, two received a increase and other two (marked with *) had a slight increase which was truncated, since Scalabrino classifier uses 16 digits score.

\begin{table}[hbt]
\centering
\scriptsize
\caption{SonarQube values after code cleaning}
\label{tab:tabelaEstatisticasSonarRefactoring}
\begin{tabular}{lrrrrrrrr}
    \toprule
    \textbf{Class} & \multicolumn{4}{c}{\textbf{Clean Code}} & \multicolumn{1}{c}{\textbf{D}} & \multicolumn{1}{c}{\textbf{L}}& \multicolumn{1}{c}{\textbf{SC}} \\
        & \textbf{C} & \textbf{I} & \textbf{A}&\textbf{R} & & \\
    \midrule
    $C_1$   &-      &-      &-      &-&62.9\%   &294    &90.7\\
    $C_2$   &\ccr-  &\ccr-  &-      &-&63.4\%   &526    &\ccg91.1\\
    $C_3$   &-      &\ccr-  &-      &-&32.1\%   &155    &93.3*\\
    $C_4$   &-      &\ccr-  &\ccr-  &-&\ccr21.0\%&\ccg586&78.9*\\
    $C_5$   &-      &\ccr-  &\ccr1  &-&\ccr9.8\%&177    &\ccg70.2\\
    $C_6$   &\ccr-  &\ccr-  &\ccr3  &-&\ccg10.3\%&\ccg825&\ccr76.5\\
    $C_7$   &-      &-      &-      &-&25.5\%   &75     &93.3\\
    $C_8$   &-      &\ccr-  &\ccr-  &-&\ccr24.6\%&\ccr589&\ccr72.3\\
    $C_9$   &-      &\ccr-  &1      &-&\ccg17.2\%&\ccg132&80.2\\
    $C_{10}$&-      &-      &-      &-&38.2\%   &48     &89.1\\
    $C_{11}$&-      &\ccr-  &\ccr5  &-&\ccr6.9\%&\ccr364&\ccr73.2\\
    $C_{12}$&-      &\ccr-  &\ccr2  &-&5.7\%    &\ccr276&\ccr67.1\\
    \bottomrule
\end{tabular}
\vspace{-5pt}
\end{table}

Only three models presented variation. The \geminiPro\ model had its results significantly more severely affected by its variation. Below we list the variation on score attribution in this scenario, with their respective SD and CV:

\begin{description}[font=\small]
    \item[\claudeSonnet]- $C_5$ 3.2 (4.27\%);
    \item[\deepSeek]- $C_6$ 1.6 (1.67\%);
    \item[\gptMini]- $C_6$ 4.8 (6.19\%) and $C_{11}$ 4.8 (5.89\%);
    \item[\geminiPro]- $C_6$ 2.3 (3.14\%), $C_7$ 1.8 (1.95\%) and $C_8$ 1.8 (2.19\%);
    \item[\llamaOito]- $C_5$ 6.3 (10.20\%), $C_7$ 1.6 (1.87\%), $C_8$ 2.4 (2.96\%), $C_{10}$ 3.0 (3.42\%) and $C_{12}$ 1.6 (1.96\%);
\end{description}

In Table \ref{tab:scoreMedioPorClasseELLMRefactoring}, we see that \ac{LLM}s showed 16 instances of score increases, highlighted in green, being six with no statistical significance, in gray color and two with moderate significance in yellow. Also, we have 14 instances of score reductions in red color, with three of these reductions with no significance in gray. \claudeSonnet\ was the only one to not present statistical noise.

\begin{table}[hbt]
    \centering
    \tiny
    \caption{Average score by class and LLM - Clean Code}
    \label{tab:scoreMedioPorClasseELLMRefactoring}
    \begin{tabular}{lrrrrrrrrrr}
        \toprule
         & \multicolumn{2}{c}{\textbf{ChatGPT}} & \multicolumn{2}{c}{\textbf{Gemini}} & \multicolumn{2}{c}{\textbf{Llama}}& \multicolumn{2}{c}{\textbf{Claude}}&\textbf{DeepSeek}&\textbf{SC}\\
        \cmidrule{2-10}
        \textbf{Class} 
                & 4o      & 4o-mini   & 2.0-Pro  & 2.0-Flash & 3.1-405B& 3.1-8B & 3.7-Sonnet & 3-Haiku&V3&\\
        \midrule
        $C_1$   & 0.0\%   & 0.0\%     & 0.0\%      & 0.0\%  & 0.0\%   &0.0\%     &\ccg5.6\%  & 0.0 &    0.0 & 0.0\% \\
        $C_2$   & 0.0\%   & 0.0\%     & 0.0\%      & 0.0\%  & 0.0\%   &\ccr-3.1\%& 0.0\%     & 0.0 &    0.0 & \ccg0.13\% \\
        $C_3$   & 0.0\%   & 0.0\%     & 0.0\%      & 0.0\%  & 0.0\%   &\ccr-7.6\%& 0.0\%     & 0.0 &    0.0 & \ccg0.03\% \\
        $C_4$   & 0.0\%   & \ccc2.4\% & \ccy9.6\%  & 0.0\%  & \ccc1.8\%&0.0\%    & 0.0\%     & 0.0 &   0.0 & \ccg0.02\% \\
        $C_5$   & 0.0\%   & 0.0\%     & \ccr-6.7\% & 0.0\%  & 0.0\%   &\ccr-22.5\%&\ccg13.8\%& 0.0 &   0.0 & \ccg8.97\% \\
        $C_6$   & 0.0\%   & \ccc4.0\% & \ccc3.5\%  & 0.0\%  & 0.0\%   &\ccg33.3\%&\ccr-11.8\%& 0.0 &    0.0 & \ccr-1.11\% \\
        $C_7$   &\ccc0.5\%& 0.0\%     &\ccr-4.7\%  & 0.0\%  & 0.0\%   &\ccg5.6\% & 0.0\%     & 0.0 &\ccc-0.5& 0.0\% \\
        $C_8$   & 0.0\%   & 0.0\%     & \ccg15.0\% & 0.0\%  & 0.0\%   &\ccg35.8\%&\ccr-11.8\%& 0.0 &    0.0 & \ccr-3.38\% \\
        $C_9$   & 0.0\%   & 0.0\%     &\ccc-2.9\%  & 0.0\%  & \ccr-5.9\%&\ccg6.2\% & 0.0\%   & 0.0 &  0.0 & 0.0\% \\
        $C_{10}$& 0.0\%   & \ccr-11.8\%&0.0\%      & 0.0\%  & 0.0\%   &\ccc1.6\% & 0.0\%     & 0.0 &    0.0 & 0.0\% \\
        $C_{11}$& 0.0\%   & \ccy9.3\% &\ccc-2.8\%  & 0.0\%  & 0.0\%   &\ccr-25.0\%& 0.0\%    & 0.0 &   0.0 & \ccr-1.10\% \\
        $C_{12}$& 0.0\%   & 0.0\%     & 0.0\%      & 0.0\%  & 0.0\%   &\ccr-5.3\%&\ccg15.4\% & 0.0 &    0.0 & \ccr-2.03\% \\
        \bottomrule
    \end{tabular}
\end{table}

For this scenario, Table \ref{tab:agreementCleanCode} shows a result equivalent to the original code. All LLMs agree with SC, marking all classes as readable.

\begin{table}[htbp]
\vspace{-5pt}
    \centering
    \scriptsize
    \caption{Agreement with Scalabrino Classifier - Clean Code}
    \begin{tabular}{lrrrrrrrrr}
        \toprule
        & \multicolumn{2}{c}{\textbf{ChatGPT}} & \multicolumn{2}{c}{\textbf{Gemini}} & \multicolumn{2}{c}{\textbf{Llama}}& \multicolumn{2}{c}{\textbf{Claude}}&\textbf{Deep}\\
        \cmidrule{2-9}
         \textbf{Metric}& 4o & 4o & 2.0 & 2.0 & 3.1 & 3.1 & 3.7 & 3&\textbf{Seek}\\
            &      & mini & Pro  & Flash& 405B & 8B & Sonnet & Haiku&V3\\
        \midrule
        Precision & 100\% & 100\% & 100\% & 100\% & 100\% & 100\% & 100\% & 100\% & 100\% \\
        Recall & 100\% & 100\% & 100\% & 100\% & 100\% & 100\% & 100\% & 100\% & 100\% \\
        F1-Score & 100\% & 100\% & 100\% & 100\% & 100\% & 100\% & 100\% & 100\% & 100\% \\
        \bottomrule
    \end{tabular}
    \label{tab:agreementCleanCode}
    \vspace{-5pt}
\end{table}

Table \ref{tab:thematic_coding} shows the results of thematic coding performed on LLMs reasoning for their scores. The table quantifies the frequency of themes mentioned by the LLMs for each of the four scenarios, based on the inductive thematic analysis methodology described in Section \ref{sec:mehodology}. The results reveal how the LLMs' qualitative assessments changed in response to each intervention:

\textbf{Original Code (OC):} This column establishes the baseline, showing high praise for "Good Code Structure" (1030 mentions) and "Good Readability" (1068 mentions). Mentions for "Good Documentation" (287) and "Good Variable/Method Names" (346) were also strong.

\textbf{Comments Removed (I1):} This intervention caused the expected sharp decline in comments about "Good Documentation" (from 287 to 32) and an increase in "Lack of Documentation/Comments" (from 17 to 64) and Unprofessional Comments (from 35 to 0). Consequently, mentions of "Good Readability" also decreased from 1068 to 933.

\textbf{Confusing Names (I2):} This scenario had the most dramatic impact on the LLMs' reasoning. Mentions of "Poor Variable/Method Names" skyrocketed from 4 to 890, while "Poor Readability" jumped from 105 in the original code to 561. This demonstrates that the LLMs' reasoning directly identified the semantic failure of the intervention, confirming their sensitivity to poor identifier choices. Also "Good Readability" plummeted from 1068 to 438, and "Good Code Structure" dropped from 1030 to 599.

\textbf{Code Smells Fixed (I3):} Fixing code smells led to a decrease in mentions of "Bad Code Structure" (from 125 to 116) and a significant increase in praise for "Good Error Handling" (from 65 to 92), suggesting these were the types of issues addressed. However, fixing smells did not uniformly improve the perception of readability, as mentions of "Poor Readability" slightly increased (from 105 to 133).

\begin{table}[hbt]
    \centering
    \footnotesize
    \caption{Code Evaluation Across Scenarios}
    \label{tab:thematic_coding}
    \begin{tabular}{lrrrr}
        \toprule
        \textbf{Theme} & \textbf{OC} & \textbf{I1} & \textbf{I2} & \textbf{I3} \\
        \midrule
        Adherence to Best Practices & 267 & 226 & 98 & 263 \\
        Good Implementation/Logic & 237 & 216 & 105 & 256 \\
        Good Use of Language Features & 78 & 88 & 15 & 89 \\
        Consistent Formatting & 182 & 178 & 53 & 208 \\
        Inconsistent Formatting & 2 & 22 & 25 & 8 \\
        Good Code Structure & 1030 & 983 & 599 & 1115 \\
        Bad Code Structure & 125 & 155 & 74 & 116 \\
        Good Documentation & 287 & 32 & 161 & 250 \\
        Lack of Documentation/Comments & 17 & 64 & 31 & 22 \\
        Unprofessional Comments & 35 & 0 & 42 & 40 \\
        Good Error Handling & 65 & 132 & 20 & 92 \\
        Poor Error Handling & 19 & 72 & 18 & 28 \\
        Good Readability & 1068 & 933 & 438 & 1021 \\
        Poor Readability & 105 & 99 & 561 & 133 \\
        Good Variable/Method Names & 346 & 457 & 111 & 390 \\
        Poor Variable/Method Names & 4 & 2 & 890 & 22 \\
        \bottomrule
        \end{tabular}
    \vspace{-5pt}
\end{table}


\section{Discussion}\label{sec:discussion}
With each intervention in the code, the \ac{LLM}s presented different results. In this section, we will discuss the results presented.

In the original code scenario, the \ac{LLM}s showed a level of correlation with \ac{SC}, except for \geminiPro\ and \gptO.

After removing comments (\(OC \rightarrow I1\)), all models showed at least one score reduction. The LLMs proved to be more lenient than the SC, correctly identifying most readable classes but at the cost of misclassifying several unreadable ones as described in Table \ref{tab:agreementNoComments}. SonarQube showed a slight increase in Code Smells, as explained in step \ref{\itemE}, and also indicated a significant drop in the percentage of comments and the total number of lines in the classes. \ac{SC} showed a consistent and strong decrease of score for all classes, except for $C_5$ that received a strong increment, those new scores resulted in two classes falling as 'neutral' classification.

As described by Good and Poor Readability on Table \ref{tab:thematic_coding}, \ac{LLM}s appear to consider that the absence of comments is not sufficient to drastically reduce readability, in most cases. There may be some other aspects of code that interact with the presence (or absence) of comments, which could interfere in the \ac{LLM}s analysis.

Changing the code to use confusing identifier names (\(OC \rightarrow I2\)) caused a significant fluctuation in scores across all models as well as an increase in the magnitude of its variability. Both Gemini models and \claudeSonnet\ showed the greatest drop in scores, being \claudeSonnet\ more consistent. All models showed some level of sensitivity to the unconventional names used in the experiment, significantly stronger comparing with reference model. But there were cases when a model did not reduce the score, as for $C_{10}$, the smallest class with 48 lines. As expected, SonarQube did not show any changes in its indicators after the identifier names were altered. All models were capable to criticize the semantic conflict between the structure of the presented algorithms and the adopted names, as mapped by thematic coding.

The \ac{SC} showed some level of sensitivity in this scenario. That could be explained by the fact that this classifier performs a correlation between the terms in comments and terms in code, measuring the overlap between both (CIC measure). The use of confusing names reduced this correlation. This algorithm also takes into account the use of full formed words present in dictionary, as we have chosen to use this type of change, this measurement was not affected. Another metric used is called Textual Coherence, that is a measurement of vocabulary overlap between all possible pairs of blocks of code \cite{scalabrino_improving_2016}. Since we have chosen to use only seasonings and vegetables names, we have created an artificial coherence, but with no semantic meaning to the code. These changes mislead \ac{SC} but not the \ac{LLM}s .

Cleaning the code (\(OC \rightarrow I3\)) resulted in all LLMs agreeing with \ac{SC}, returning to the same results of original code state. The thematic coding demonstrated that LLMs reacted to changes reducing comments about "Bad Code Structure", incrementing comments on "Good Error Handling", but with a small increase of comments about "Poor Readability".

To evaluate \ac{LLM}s variability we used 12 classes, 9 \ac{LLM} and 4 scenarios, resulting in 432 combinations and 4,320 \ac{LLM} executions. All \ac{LLM} presented variation at least in one scenario. 47 (10.88\%) combinations out of 432 presented variation, \geminiPro\ was responsible by 18 (38.30\%) of all variations occurrence, being the only to present variation in all four scenarios. Conversely, \geminiFlash\ was the only one to present no variation throughout all executions. 

31 (65.96\%) variations were equal to or lower than 5\% and scenario I2 (confusing names) had 9 variations above 5\% followed by I3 (code cleaning) with 3. 
Near half of variations did not compromise the statistical relevance of emitted scores. Therefore most of \ac{LLM} score (94.02\%) does not present statistical noise. 

The reference model uses textual features to compose its readability score\cite{scalabrino_improving_2016}, but \ac{LLM}s have shown the ability to evaluate semantic aspects of texts present in code, revealing deeper sensitivity to identifiers name changes, not present in the reference model. This was shown in scenario I2. 

The same characteristics seem to affect the I1 scenario. The reference model has shown to be more sensitive to comments removal, mostly due to its metrics that correlate code with comments. Conversely, the \ac{LLM}s seemed to be capable to extract the easiness of code to be read, even without comments. This fact raises a question: does \ac{LLM} have used also the code semantics to understand it, dispensing documentation for algorithm comprehension?

\section{Limitations and threats to validity}\label{sec:limitationsAndFutureWork}

\textcite{shadish_experimental_2001} enlist several types of threats to validity common to quasi-experiment design. The main categories are: \textit{Construct Validity, Internal Validity, External Validity} and \textit{Statistical Conclusion Validity}.
In this section we detail all mapped threats to validity, limitations and future works derived from this study, grouped by the above listed categories.

\subsubsection*{Construct Validity}

Source code readability is a complex construct to be explained and characterized. Failing to do so, could lead us to confound it with other aspects of source code quality as well as incorrect inference about the the correlations observed from every operation (intervention) into the source code. To mitigate this threat we performed an extensive literature review on the main related studies and adopted a recognized state of the art model obtained in a state of use that required no implementation or intervention to work. Even so it is know that readability models have a limited capability to assess code readability the same way humans do \cite{fakhoury_improving_2019, sergeyuk_reassessing_2024}, being this a existent threat to validity.

The adopted strategy was also a mitigating factor for other two threats: The mono-operation bias caused by our strategy to perform a single stereotyped operation at a time (i.e removing comments, replacing identifiers names), which under represent the broad concept of source code readability. And the mono-method bias caused by the self produced source code interventions, which could lead to a bias of interventions performed based on what the research team understands to be a \textit{"code readability intervention"}.

There is a threat to confounding the constructs with the levels of constructs. In this quai-experiment, we performed interventions that were limited to a very specific set of characteristics related to source code readability. We are not able to generalize it to the broad concept of readability. The findings we present can lead us to conclusions related solely to those manipulated characteristics. The control mechanisms used guarantee that the interventions achieved the goal to change the source code readability, for those aspects solely.

\subsubsection*{Internal Validity}

We tried to select algorithms with different levels of readability. Since it was an arbitrary selection, we are subjected to the selection bias. To remediate any undesirable effect from this threat, we executed a pretest to all selected classes using all selected \ac{LLM}s, comparing the results with the control measurement tools.

We also recognize that we are subject to the internal validity threat described by \textcite{shadish_experimental_2001} as the \textit{"Instrumentation Threat"}. Since the \ac{LLM}s were used through web API, we have no control over its internal state and runtime configuration. The providers could change any parameter, causing a difference in its behavior which could be confused as a difference caused by code interventions. To prevent that, we have executed all API calls to the same \ac{LLM} into the same day, in a short time interval.

\subsubsection*{External Validity}
The source code used in this work is written in Java, as well as one of the control tools used (\ac{SC}), which was able to perform readability analysis for Java Code. This is a known threat to validity of our findings, as the interventions executed could have different outcomes for different programming languages. Despite the fact the Java is the most studied language in this subject, it is important to reproduce the study with different programming languages in the future.

We have chosen not to study the interactions of causal relationship over code interventions (i.e treatment variations); this is a known limitation of the collected results. The \ac{SC} model, for example, measures the relationship between algorithm terms and documentation terminology. By performing the interventions in isolation we introduced a bias to the results, as detailed in section \ref{sec:discussion}. As the goal of this work is to characterize how \ac{LLM}s respond to such isolated changes, this approach was considered enough to answer the research questions. In the future a study could be performed to analyze how the interaction between different code interventions affects the perception of readability.

Our work uses the results of Scalabrino's Classifier (SCO) as ground truth, despite evidence that it can be unreliable under certain conditions \cite{piantadosi_how_2020}, and that it does not fully capture code readability as perceived by humans \cite{pantiuchina_improving_2018, piantadosi_how_2020, sergeyuk_reassessing_2024}. This is a known threat to validity of this work and a limitation. The results only prove the comparability of LLMs with SCO, even though they produce results semantically distinct.

\subsubsection*{Statistical Conclusion Validity}

Since we selected a small set of algorithms to perform manual and controlled changes, we are subject to a threat of having low statistical power in our results as well as other statistical threats, such as inaccurate results. To ensure the results described are adequate, we performed complementary statistical tests, using non parametric tests. With these tests we were able to detect non statistically relevant results, and highlight them during results description. Still, the low amount of code used poses a threat to the generalizability of the results obtained.

An extraneous variance in the experimental setting is a threat caused by the known variance of \ac{LLM}s answers to a prompt. To mitigate this threat we performed a 10 fold execution for every evaluation and executed statistical analysis for the results. This approach allowed us to draw the conclusions on when a \ac{LLM} score have changes because of a code intervention and when its just random noise.

The unreliability of treatment implementation is a threat that could raise from the manual interventions done into the code. In order to mitigate it, we decided to execute very simple interventions, as described in section \ref{sec:\itemD}, just removing the comments, replacing identifiers name by a dictionary of words, and performing SonarQube's guided fixing for Code Smells. With this approach we have guaranteed a strict and reproducible method for manual code intervention.

\vspace{-5pt}
\section{Conclusion and future work}\label{sec:conclusion}
This work explored the sensitivity of \ac{LLM} models to interventions related to source code readability. Our results demonstrated that all models exhibited some level of reaction to the changes. However, the results also raised some new questions.

(1) The results of this research are promising  regarding the possibility of using LLMs to generate a metric that assesses code quality attributes related to readability. However, when comparing the results demonstrated by \ac{LLM}s with the reference model, we observe differences in their scores. 

The \ac{LLM} analysis did not fully agree with the reference model, but behaved differently. It was influenced by the changes performed on source code; but this resulted in a score statistically different from the reference model in most cases. As future work, we plan to compare the \ac{LLM}s results with human evaluation.

(2) With respect to the attributes that affect readability, we observed that \ac{LLM}s show more sensitivity to the changes made to identifiers name (I2), when intentionally confusing names have been used. This behavior raised a hypothesis that the \ac{LLM}s are relying on semantic meaning of words to evaluate code quality.

When comments were removed (I1), it was expected to see a reduction in readability scores. Although there was some reduction (20.83\% of subjects), the majority of scores remained unchanged, with even a few improvements. This behavior could also be caused by the reliance on semantic meaning, but now extracted from the algorithm itself. There may be other code attributes that could be influencing \ac{LLM}s to generate a score for code readability. This should be investigated in future work.

The clean code scenario (I3) demonstrated disappointing results, with more score reductions (9) than improvements (5), a lot of statistical noise (11) and mostly no change (71). Even for the control tool \ac{SC} there was 4 improvements, 4 reductions and 4 classes with no change. We conclude that fixing code smells does not guarantee an improvement in code readability.

(3) In relation to the inherent variability of \ac{LLM}s, our results show that while this variability affects the assessment of code quality, its occurrence rate is low. In the four scenarios evaluated, there were 9.37\% (OC and I1), 12.50\% (I3) and 14.58\% (I2) of combinations with standard deviation different from zero. Most of them affected the statistical reliability of results, with exception of scenario I2. We recommend future evaluations of \ac{LLM}s based on this study, to perform additional redundant tests, followed by statistical analysis, to check the validity of obtained results. A ten round execution demonstrated to be sufficient to provide data for a statistical validity check.

Overall, this study demonstrates that \ac{LLM}s have the potential to be used as a complementary technique for evaluating quality aspects involving semantics, such as identifier names, comments content, and code documentation. Further research is needed to investigate the comparability of the metrics generated by \ac{LLM}s with human developers' perceptions and to develop strategies to address the variability in the model's responses and their activation costs.

\vspace{-5pt}
\section*{Data Availability}

The source code of the programs used to execute the LLMs, as well as the source codes of the analyzed programs and the datasets from these analyses, are available in the GitHub repository referenced in this article \cite{LLMSonarQuarkusAnalysis}.


\printbibliography

@misc{LLMSonarQuarkusAnalysis,
  author = {Regis, Igor},
  title = {LLMSonarQuarkusAnalysis},
  year = {2024},
  publisher = {GitHub},
  journal = {GitHub repository},
  howpublished = {\url{https://github.com/igorregis/LLMSonarQuarkusAnalysis}}
}

@inproceedings{buse_metric_2008,
    address = {Seattle WA USA},
    title = {A metric for software readability},
    isbn = {978-1-60558-050-0},
    url = {https://dl.acm.org/doi/10.1145/1390630.1390647},
    doi = {10.1145/1390630.1390647},
    language = {en},
    urldate = {2025-01-27},
    booktitle = {Proceedings of the 2008 international symposium on {Software} testing and analysis},
    publisher = {ACM},
    author = {Buse, Raymond P.L. and Weimer, Westley R.},
    month = jul,
    year = {2008},
    pages = {121--130},
}

@inproceedings{scalabrino_automatically_2017,
    address = {Urbana, IL},
    title = {Automatically assessing code understandability: {How} far are we?},
    isbn = {978-1-5386-2684-9},
    shorttitle = {Automatically assessing code understandability},
    url = {http://ieeexplore.ieee.org/document/8115654/},
    doi = {10.1109/ASE.2017.8115654},
    urldate = {2024-08-08},
    booktitle = {2017 32nd {IEEE}/{ACM} {International} {Conference} on {Automated} {Software} {Engineering} ({ASE})},
    publisher = {IEEE},
    author = {Scalabrino, Simone and Bavota, Gabriele and Vendome, Christopher and Linares-Vasquez, Mario and Poshyvanyk, Denys and Oliveto, Rocco},
    month = oct,
    year = {2017},
    pages = {417--427},
}

@article{hou_large_2024,
	title = {Large {Language} {Models} for {Software} {Engineering}: {A} {Systematic} {Literature} {Review}},
	volume = {33},
	issn = {1049-331X, 1557-7392},
	shorttitle = {Large {Language} {Models} for {Software} {Engineering}},
	doi = {10.1145/3695988},
	language = {en},
	number = {8},
	urldate = {2025-05-29},
	journal = {ACM Transactions on Software Engineering and Methodology},
	author = {Hou, Xinyi and Zhao, Yanjie and Liu, Yue and Yang, Zhou and Wang, Kailong and Li, Li and Luo, Xiapu and Lo, David and Grundy, John and Wang, Haoyu},
	month = nov,
	year = {2024},
	pages = {1--79},
}

@article{wong_natural_2023,
	title = {Natural {Language} {Generation} and {Understanding} of {Big} {Code} for {AI}-{Assisted} {Programming}: {A} {Review}},
	volume = {25},
	copyright = {https://creativecommons.org/licenses/by/4.0/},
	issn = {1099-4300},
	shorttitle = {Natural {Language} {Generation} and {Understanding} of {Big} {Code} for {AI}-{Assisted} {Programming}},
	doi = {10.3390/e25060888},
	language = {en},
	number = {6},
	journal = {Entropy},
	author = {Wong, Man-Fai and Guo, Shangxin and Hang, Ching-Nam and Ho, Siu-Wai and Tan, Chee-Wei},
	month = jun,
	year = {2023},
	pages = {888},
}

@article{scalabrino_automatically_2019,
	title = {Automatically {Assessing} {Code} {Understandability}},
	volume = {47},
	copyright = {https://ieeexplore.ieee.org/Xplorehelp/downloads/license-information/IEEE.html},
	issn = {0098-5589, 1939-3520, 2326-3881},
	doi = {10.1109/TSE.2019.2901468},
	number = {3},
	journal = {IEEE Transactions on Software Engineering},
	author = {Scalabrino, Simone and Bavota, Gabriele and Vendome, Christopher and Linares-Vásquez, Mario and Poshyvanyk, Denys and Oliveto, Rocco},
	month = mar,
	year = {2019},
	pages = {595--613},
}

@inproceedings{mi_enhanced_2022,
	title = {An {Enhanced} {Data} {Augmentation} {Approach} to {Support} {Multi}-{Class} {Code} {Readability} {Classification}},
	url = {http://ksiresearchorg.ipage.com/seke/seke22paper/paper130.pdf},
	doi = {10.18293/SEKE2022-130},
	author = {Mi, Qing and Hao, Yiqun and Wu, Maran and Ou, Liwei},
	month = jul,
	year = {2022},
	pages = {48--53},
}

@book{wohlin_experimentation_2012,
	address = {Berlin, Heidelberg},
	title = {Experimentation in {Software} {Engineering}},
	copyright = {http://www.springer.com/tdm},
	isbn = {978-3-642-29043-5 978-3-642-29044-2},
	url = {http://link.springer.com/10.1007/978-3-642-29044-2},
	language = {en},
	publisher = {Springer Berlin Heidelberg},
	author = {Wohlin, Claes and Runeson, Per and Höst, Martin and Ohlsson, Magnus C. and Regnell, Björn and Wesslén, Anders},
	year = {2012},
	doi = {10.1007/978-3-642-29044-2},
}

@inproceedings{pantiuchina_improving_2018,
	address = {Madrid},
	title = {Improving {Code}: {The} ({Mis}) {Perception} of {Quality} {Metrics}},
	isbn = {978-1-5386-7870-1},
	shorttitle = {Improving {Code}},
	url = {https://ieeexplore.ieee.org/document/8530019/},
	doi = {10.1109/ICSME.2018.00017},
	booktitle = {2018 {IEEE} {International} {Conference} on {Software} {Maintenance} and {Evolution} ({ICSME})},
	publisher = {IEEE},
	author = {Pantiuchina, Jevgenija and Lanza, Michele and Bavota, Gabriele},
	month = sep,
	year = {2018},
	pages = {80--91},
}

@book{guest_applied_2012,
	address = {2455 Teller Road, Thousand Oaks California 91320 United States},
	title = {Applied {Thematic} {Analysis}},
	isbn = {978-1-4129-7167-6 978-1-4833-8443-6},
	publisher = {SAGE Publications, Inc.},
	author = {Guest, Greg and MacQueen, Kathleen and Namey, Emily},
	year = {2012},
	doi = {10.4135/9781483384436},
}

@inproceedings{hu_how_2024,
	address = {Sacramento CA USA},
	title = {How {Effectively} {Do} {Code} {Language} {Models} {Understand} {Poor}-{Readability} {Code}?},
	isbn = {979-8-4007-1248-7},
	url = {https://dl.acm.org/doi/10.1145/3691620.3695072},
	doi = {10.1145/3691620.3695072},
	language = {en},
	urldate = {2025-05-31},
	booktitle = {Proceedings of the 39th {IEEE}/{ACM} {International} {Conference} on {Automated} {Software} {Engineering}},
	publisher = {ACM},
	author = {Hu, Chao and Chai, Yitian and Zhou, Hao and Meng, Fandong and Zhou, Jie and Gu, Xiaodong},
	month = oct,
	year = {2024},
	pages = {795--806},
}

@inproceedings{sergeyuk_reassessing_2024,
	address = {Lisbon Portugal},
	title = {Reassessing {Java} {Code} {Readability} {Models} with a {Human}-{Centered} {Approach}},
	isbn = {979-8-4007-0586-1},
	url = {https://dl.acm.org/doi/10.1145/3643916.3644435},
	doi = {10.1145/3643916.3644435},
	language = {en},
	urldate = {2025-05-26},
	booktitle = {Proceedings of the 32nd {IEEE}/{ACM} {International} {Conference} on {Program} {Comprehension}},
	publisher = {ACM},
	author = {Sergeyuk, Agnia and Lvova, Olga and Titov, Sergey and Serova, Anastasiia and Bagirov, Farid and Kirillova, Evgeniia and Bryksin, Timofey},
	month = apr,
	year = {2024},
	pages = {225--235},
}

@inproceedings{raymond_reading_1991,
	address = {Toronto, Ontario, Canada},
	series = {{CASCON} '91},
	title = {Reading source code},
	urldate = {2025-05-02},
	booktitle = {Proceedings of the 1991 conference of the {Centre} for {Advanced} {Studies} on {Collaborative} research},
	publisher = {IBM Press},
	author = {Raymond, Darrell R.},
	month = oct,
	year = {1991},
	pages = {3--16},
}

@article{dorn_general_2012,
	title = {A general software readability model},
	volume = {5},
	url = {http://www.cs.virginia.edu/weimer/students/dorn-mcs-paper.pdf},
	journal = {MCS Thesis},
	author = {Dorn, Jonathan},
	year = {2012},
	pages = {11--14},
}

@inproceedings{white_prompt_2023,
	address = {Monticello, Illinois, USA},
	title = {A {Prompt} {Pattern} {Catalog} to {Enhance} {Prompt} {Engineering} with {ChatGPT}},
	url = {https://www.dre.vanderbilt.edu/~schmidt/PDF/PLoP-patterns.pdf},
	author = {White, Jules and Fu, Quchen and Hays, Sam and Sandborn, Michael and Olea, Carlos and Gilbert, Henry and Elnashar, Ashraf and Spencer-Smith, Jesse and Schmidt, Douglas C.},
	month = oct,
	year = {2023},
}

@article{obrien_expectation_2004,
	title = {Expectation based, inference‐based, and bottom‐up software comprehension},
	volume = {16},
	copyright = {http://onlinelibrary.wiley.com/termsAndConditions\#vor},
	issn = {1532-060X, 1532-0618},
	url = {https://onlinelibrary.wiley.com/doi/10.1002/smr.307},
	doi = {10.1002/smr.307},
	language = {en},
	number = {6},
	journal = {Journal of Software Maintenance and Evolution: Research and Practice},
	author = {O'Brien, Michael P. and Buckley, Jim and Shaft, Teresa M.},
	month = nov,
	year = {2004},
	pages = {427--447},
}

@misc{vitale_personalized_2025,
	title = {Personalized {Code} {Readability} {Assessment}: {Are} {We} {There} {Yet}?},
	shorttitle = {Personalized {Code} {Readability} {Assessment}},
	url = {http://arxiv.org/abs/2503.07870},
	doi = {10.48550/arXiv.2503.07870},
	urldate = {2025-03-23},
	publisher = {arXiv},
	author = {Vitale, Antonio and Guglielmi, Emanuela and Oliveto, Rocco and Scalabrino, Simone},
	month = mar,
	year = {2025},
	note = {arXiv:2503.07870 [cs]},
}

@inproceedings{cheng_what_2022,
	address = {Singapore Singapore},
	title = {What improves developer productivity at google? code quality},
	isbn = {978-1-4503-9413-0},
	shorttitle = {What improves developer productivity at google?},
	url = {https://dl.acm.org/doi/10.1145/3540250.3558940},
	doi = {10.1145/3540250.3558940},
	language = {en},
	booktitle = {Proceedings of the 30th {ACM} {Joint} {European} {Software} {Engineering} {Conference} and {Symposium} on the {Foundations} of {Software} {Engineering}},
	publisher = {ACM},
	author = {Cheng, Lan and Murphy-Hill, Emerson and Canning, Mark and Jaspan, Ciera and Green, Collin and Knight, Andrea and Zhang, Nan and Kammer, Elizabeth},
	month = nov,
	year = {2022},
	pages = {1302--1313},
}

@inproceedings{buse_information_2012,
	address = {Zurich},
	title = {Information needs for software development analytics},
	isbn = {978-1-4673-1066-6 978-1-4673-1067-3},
	url = {http://ieeexplore.ieee.org/document/6227122/},
	doi = {10.1109/ICSE.2012.6227122},
	booktitle = {2012 34th {International} {Conference} on {Software} {Engineering} ({ICSE})},
	publisher = {IEEE},
	author = {Buse, Raymond P. L. and Zimmermann, Thomas},
	month = jun,
	year = {2012},
	pages = {987--996},
}

@misc{wang_large_2023,
	title = {Large {Language} {Models} {Are} {Zero}-{Shot} {Text} {Classifiers}},
	url = {http://arxiv.org/abs/2312.01044},
	publisher = {arXiv},
	author = {Wang, Zhiqiang and Pang, Yiran and Lin, Yanbin},
	month = dec,
	year = {2023},
	note = {arXiv:2312.01044 [cs]},
}

@misc{wei_emergent_2022,
	title = {Emergent {Abilities} of {Large} {Language} {Models}},
	url = {http://arxiv.org/abs/2206.07682},
	publisher = {arXiv},
	author = {Wei, Jason and Tay, Yi and Bommasani, Rishi and Raffel, Colin and Zoph, Barret and Borgeaud, Sebastian and Yogatama, Dani and Bosma, Maarten and Zhou, Denny and Metzler, Donald and Chi, Ed H. and Hashimoto, Tatsunori and Vinyals, Oriol and Liang, Percy and Dean, Jeff and Fedus, William},
	month = oct,
	year = {2022},
	note = {arXiv:2206.07682 [cs]},
}

@misc{zhao_understanding_2023,
	title = {Understanding {Programs} by {Exploiting} ({Fuzzing}) {Test} {Cases}},
	copyright = {arXiv.org perpetual, non-exclusive license},
	url = {https://arxiv.org/abs/2305.13592},
	doi = {10.48550/ARXIV.2305.13592},
	publisher = {arXiv},
	author = {Zhao, Jianyu and Rong, Yuyang and Guo, Yiwen and He, Yifeng and Chen, Hao},
	year = {2023},
	note = {Version Number: 2},
}

@inproceedings{al_madi_novice_2021,
	address = {Madrid, Spain},
	title = {From {Novice} to {Expert}: {Analysis} of {Token} {Level} {Effects} in a {Longitudinal} {Eye} {Tracking} {Study}},
	copyright = {https://doi.org/10.15223/policy-029},
	shorttitle = {From {Novice} to {Expert}},
	url = {https://ieeexplore.ieee.org/document/9462965/},
	doi = {10.1109/ICPC52881.2021.00025},
	booktitle = {2021 {IEEE}/{ACM} 29th {International} {Conference} on {Program} {Comprehension} ({ICPC})},
	publisher = {IEEE},
	author = {Al Madi, Naser and Peterson, Cole S. and Sharif, Bonita and Maletic, Jonathan I.},
	month = may,
	year = {2021},
	pages = {172--183},
}

@article{chang_survey_2024,
	title = {A {Survey} on {Evaluation} of {Large} {Language} {Models}},
	volume = {15},
	url = {https://dl.acm.org/doi/10.1145/3641289},
	doi = {10.1145/3641289},
	language = {en},
	number = {3},
	journal = {ACM Transactions on Intelligent Systems and Technology},
	author = {Chang, Yupeng and Wang, Xu and Wang, Jindong and Wu, Yuan and Yang, Linyi and Zhu, Kaijie and Chen, Hao and Yi, Xiaoyuan and Wang, Cunxiang and Wang, Yidong and Ye, Wei and Zhang, Yue and Chang, Yi and Yu, Philip S. and Yang, Qiang and Xie, Xing},
	month = jun,
	year = {2024},
	pages = {1--45},
}

@article{scalabrino_comprehensive_2018,
	title = {A comprehensive model for code readability},
	volume = {30},
	url = {https://onlinelibrary.wiley.com/doi/10.1002/smr.1958},
	doi = {10.1002/smr.1958},
	language = {en},
	number = {6},
	journal = {Journal of Software: Evolution and Process},
	author = {Scalabrino, Simone and Linares‐Vásquez, Mario and Oliveto, Rocco and Poshyvanyk, Denys},
	month = jun,
	year = {2018},
	pages = {e1958},
}

@article{buse_learning_2010,
	title = {Learning a {Metric} for {Code} {Readability}},
	volume = {36},
	copyright = {https://ieeexplore.ieee.org/Xplorehelp/downloads/license-information/IEEE.html},
	url = {http://ieeexplore.ieee.org/document/5332232/},
	doi = {10.1109/TSE.2009.70},
	number = {4},
	journal = {IEEE Transactions on Software Engineering},
	author = {Buse, Raymond P L and Weimer, Westley R},
	month = jul,
	year = {2010},
	pages = {546--558},
}

@inproceedings{minelli_i_2015,
	address = {Florence, Italy},
	title = {I {Know} {What} {You} {Did} {Last} {Summer} - {An} {Investigation} of {How} {Developers} {Spend} {Their} {Time}},
	url = {http://ieeexplore.ieee.org/document/7181430/},
	doi = {10.1109/ICPC.2015.12},
	booktitle = {2015 {IEEE} 23rd {International} {Conference} on {Program} {Comprehension}},
	publisher = {IEEE},
	author = {Minelli, Roberto and Mocci, Andrea and Lanza, Michele},
	month = may,
	year = {2015},
	pages = {25--35},
}

@inproceedings{trockman_automatically_2018,
	address = {Gothenburg Sweden},
	title = {"{Automatically} assessing code understandability" reanalyzed: combined metrics matter},
	shorttitle = {"{Automatically} assessing code understandability" reanalyzed},
	url = {https://dl.acm.org/doi/10.1145/3196398.3196441},
	doi = {10.1145/3196398.3196441},
	language = {en},
	booktitle = {Proceedings of the 15th {International} {Conference} on {Mining} {Software} {Repositories}},
	publisher = {ACM},
	author = {Trockman, Asher and Cates, Keenen and Mozina, Mark and Nguyen, Tuan and Kästner, Christian and Vasilescu, Bogdan},
	month = may,
	year = {2018},
	pages = {314--318},
}

@inproceedings{dantas_readability_2021,
	address = {Brasil},
	title = {Readability and {Understandability} {Scores} for {Snippet} {Assessment}: an {Exploratory} {Study}},
	shorttitle = {Readability and {Understandability} {Scores} for {Snippet} {Assessment}},
	url = {https://sol.sbc.org.br/index.php/vem/article/view/17217},
	doi = {10.5753/vem.2021.17217},
	booktitle = {Anais do {IX} {Workshop} de {Visualização}, {Evolução} e {Manutenção} de {Software} ({VEM} 2021)},
	publisher = {Sociedade Brasileira de Computação - SBC},
	author = {Dantas, Carlos Eduardo C. and Maia, Marcelo A.},
	month = sep,
	year = {2021},
	pages = {46--50},
}

@inproceedings{fakhoury_improving_2019,
	address = {Montreal, QC, Canada},
	title = {Improving {Source} {Code} {Readability}: {Theory} and {Practice}},
	copyright = {https://ieeexplore.ieee.org/Xplorehelp/downloads/license-information/IEEE.html},
	isbn = {978-1-7281-1519-1},
	shorttitle = {Improving {Source} {Code} {Readability}},
	url = {https://ieeexplore.ieee.org/document/8813254/},
	doi = {10.1109/ICPC.2019.00014},
	booktitle = {2019 {IEEE}/{ACM} 27th {International} {Conference} on {Program} {Comprehension} ({ICPC})},
	publisher = {IEEE},
	author = {Fakhoury, Sarah and Roy, Devjeet and Hassan, Adnan and Arnaoudova, Vernera},
	month = may,
	year = {2019},
	pages = {2--12},
}

@inproceedings{scalabrino_improving_2016,
	address = {Austin, TX, USA},
	title = {Improving code readability models with textual features},
	isbn = {978-1-5090-1428-6},
	url = {http://ieeexplore.ieee.org/document/7503707/},
	doi = {10.1109/ICPC.2016.7503707},
	booktitle = {2016 {IEEE} 24th {International} {Conference} on {Program} {Comprehension} ({ICPC})},
	publisher = {IEEE},
	author = {Scalabrino, Simone and Linares-Vasquez, Mario and Poshyvanyk, Denys and Oliveto, Rocco},
	month = may,
	year = {2016},
	pages = {1--10},
}

@article{piantadosi_how_2020,
	title = {How does code readability change during software evolution?},
	volume = {25},
	issn = {1382-3256, 1573-7616},
	url = {https://link.springer.com/10.1007/s10664-020-09886-9},
	doi = {10.1007/s10664-020-09886-9},
	language = {en},
	number = {6},
	journal = {Empirical Software Engineering},
	author = {Piantadosi, Valentina and Fierro, Fabiana and Scalabrino, Simone and Serebrenik, Alexander and Oliveto, Rocco},
	month = nov,
	year = {2020},
	pages = {5374--5412},
}

@article{wyrich_40_2024,
	title = {40 {Years} of {Designing} {Code} {Comprehension} {Experiments}: {A} {Systematic} {Mapping} {Study}},
	volume = {56},
	issn = {0360-0300, 1557-7341},
	shorttitle = {40 {Years} of {Designing} {Code} {Comprehension} {Experiments}},
	url = {https://dl.acm.org/doi/10.1145/3626522},
	doi = {10.1145/3626522},
	language = {en},
	number = {4},
	journal = {ACM Computing Surveys},
	author = {Wyrich, Marvin and Bogner, Justus and Wagner, Stefan},
	month = apr,
	year = {2024},
	pages = {1--42},
}

@misc{shen_benchmarking_2022,
	title = {Benchmarking {Language} {Models} for {Code} {Syntax} {Understanding}},
	copyright = {Creative Commons Attribution 4.0 International},
	url = {https://arxiv.org/abs/2210.14473},
	doi = {10.48550/ARXIV.2210.14473},
	publisher = {arXiv},
	author = {Shen, Da and Chen, Xinyun and Wang, Chenguang and Sen, Koushik and Song, Dawn},
	year = {2022},
	note = {Version Number: 1},
}

@article{mi_improving_2018,
	title = {Improving code readability classification using convolutional neural networks},
	volume = {104},
	issn = {09505849},
	url = {https://linkinghub.elsevier.com/retrieve/pii/S0950584918301496},
	doi = {10.1016/j.infsof.2018.07.006},
	language = {en},
	journal = {Information and Software Technology},
	author = {Mi, Qing and Keung, Jacky and Xiao, Yan and Mensah, Solomon and Gao, Yujin},
	month = dec,
	year = {2018},
	pages = {60--71},
}

@article{mi_towards_2022,
	title = {Towards using visual, semantic and structural features to improve code readability classification},
	volume = {193},
	issn = {01641212},
	url = {https://linkinghub.elsevier.com/retrieve/pii/S0164121222001467},
	doi = {10.1016/j.jss.2022.111454},
	language = {en},
	journal = {Journal of Systems and Software},
	author = {Mi, Qing and Hao, Yiqun and Ou, Liwei and Ma, Wei},
	month = nov,
	year = {2022},
	pages = {111454},
}

@inproceedings{schankin_descriptive_2018,
	address = {Gothenburg Sweden},
	title = {Descriptive compound identifier names improve source code comprehension},
	isbn = {978-1-4503-5714-2},
	url = {https://dl.acm.org/doi/10.1145/3196321.3196332},
	doi = {10.1145/3196321.3196332},
	language = {en},
	booktitle = {Proceedings of the 26th {Conference} on {Program} {Comprehension}},
	publisher = {ACM},
	author = {Schankin, Andrea and Berger, Annika and Holt, Daniel V. and Hofmeister, Johannes C. and Riedel, Till and Beigl, Michael},
	month = may,
	year = {2018},
	pages = {31--40},
}

@inproceedings{manh_vault_2023,
	address = {Singapore},
	title = {The {Vault}: {A} {Comprehensive} {Multilingual} {Dataset} for {Advancing} {Code} {Understanding} and {Generation}},
	shorttitle = {The {Vault}},
	url = {https://aclanthology.org/2023.nlposs-1.25.pdf},
	booktitle = {Proceedings of the 3rd {Workshop} for {Natural} {Language} {Processing} {Open} {Source} {Software} ({NLP}-{OSS} 2023)},
	publisher = {ACL Association for Computational Linguistics},
	author = {Manh, Dung Nguyen and Hai, Nam Le and Dau, Anh T. V. and Nguyen, Anh Minh and Nghiem, Khanh and Guo, Jin and Bui, Nghi D. Q.},
	month = dec,
	year = {2023},
}

@inproceedings{wang_codet5_2023,
	title = {{CodeT5}+: {Open} {Code} {Large} {Language} {Models} for {Code} {Understanding} and {Generation}},
	shorttitle = {{CodeT5}+},
	url = {https://aclanthology.org/2023.emnlp-main.68/},
	doi = {10.18653/v1/2023.emnlp-main.68},
	booktitle = {Proceedings of the 2023 {Conference} on {Empirical} {Methods} in {Natural} {Language} {Processing}},
	publisher = {Association for Computational Linguistics},
	author = {Wang, Yue and Le, Hung and Gotmare, Akhilesh Deepak and Bui, Nghi D. Q. and Li, Junnan and Hoi, Steven C. H.},
	month = dec,
	year = {2023},
}

@inproceedings{mi_inception_2018,
	address = {Christchurch New Zealand},
	title = {An {Inception} {Architecture}-{Based} {Model} for {Improving} {Code} {Readability} {Classification}},
	isbn = {978-1-4503-6403-4},
	url = {https://dl.acm.org/doi/10.1145/3210459.3210473},
	doi = {10.1145/3210459.3210473},
	language = {en},
	urldate = {2025-02-15},
	booktitle = {Proceedings of the 22nd {International} {Conference} on {Evaluation} and {Assessment} in {Software} {Engineering} 2018},
	publisher = {ACM},
	author = {Mi, Qing and Keung, Jacky and Xiao, Yan and Mensah, Solomon and Mei, Xiupei},
	month = jun,
	year = {2018},
	pages = {139--144},
}

@book{shadish_experimental_2001,
	address = {Belmont, CA},
	edition = {Nachdr.},
	title = {Experimental and quasi-experimental designs for generalized causal inference},
	isbn = {978-0-395-61556-0},
	language = {eng},
	publisher = {Wadsworth Cengage Learning},
	author = {Shadish, William R. and Cook, Thomas D. and Campbell, Donald T.},
	year = {2001},
}

@inproceedings{simoes_evaluating_2024,
	address = {Salvador Bahia Brazil},
	title = {Evaluating {Source} {Code} {Quality} with {Large} {Languagem} {Models}: a comparative study},
	isbn = {979-8-4007-1777-2},
	shorttitle = {Evaluating {Source} {Code} {Quality} with {Large} {Languagem} {Models}},
	url = {https://dl.acm.org/doi/10.1145/3701625.3701650},
	doi = {10.1145/3701625.3701650},
	language = {en},
	urldate = {2025-01-01},
	booktitle = {Proceedings of the {XXIII} {Brazilian} {Symposium} on {Software} {Quality}},
	publisher = {ACM},
	author = {Simões, Igor Regis Da Silva and Venson, Elaine},
	month = nov,
	year = {2024},
	pages = {103--113},
}

@inproceedings{kanade_learning_2020,
	series = {Proceedings of {Machine} {Learning} {Research}},
	title = {Learning and {Evaluating} {Contextual} {Embedding} of {Source} {Code}},
	volume = {119},
	url = {https://proceedings.mlr.press/v119/kanade20a.html},
	booktitle = {Proceedings of the 37th {International} {Conference} on {Machine} {Learning}},
	publisher = {PMLR},
	author = {Kanade, Aditya and Maniatis, Petros and Balakrishnan, Gogul and Shi, Kensen},
	editor = {III, Hal Daumé and Singh, Aarti},
	month = jul,
	year = {2020},
	pages = {5110--5121},
}

@book{kruchten_managing_2019,
	address = {Sydney},
	title = {Managing {Technical} {Debt} {Reducing} {Friction} in {Software} {Development}},
	isbn = {978-0-13-564596-3},
	language = {eng},
	publisher = {Pearson Education, Limited},
	author = {Kruchten, Philippe and Ozkaya, Ipek},
	year = {2019},
	note = {OCLC: 1338840518},
}

@article{besker_influence_2020,
	title = {The influence of {Technical} {Debt} on software developer morale},
	volume = {167},
	issn = {01641212},
	url = {https://linkinghub.elsevier.com/retrieve/pii/S0164121220300674},
	doi = {10.1016/j.jss.2020.110586},
	language = {en},
	urldate = {2024-08-09},
	journal = {Journal of Systems and Software},
	author = {Besker, Terese and Ghanbari, Hadi and Martini, Antonio and Bosch, Jan},
	month = sep,
	year = {2020},
	pages = {110586},
}

@inproceedings{boehm_quantitative_1976,
	address = {Washington, DC, USA},
	series = {{ICSE} '76},
	title = {Quantitative evaluation of software quality},
	booktitle = {Proceedings of the 2nd {International} {Conference} on {Software} {Engineering}},
	publisher = {IEEE Computer Society Press},
	author = {Boehm, B. W. and Brown, J. R. and Lipow, M.},
	year = {1976},
	note = {event-place: San Francisco, California, USA},
	pages = {592--605},
}

@phdthesis{bexell_software_2020,
	address = {Karlskrona, Sweden},
	title = {Software {Source} {Code} {Readability}: {A} {Mapping} {Study}},
	url = {https://www.diva-portal.org/smash/record.jsf?pid=diva2%3A1452612&dswid=-8633},
	school = {Blekinge Institute of Technology},
	author = {Bexell, Andreas},
	month = aug,
	year = {2020},
}

@book{fowler_refactoring_1999,
	address = {Reading, MA},
	series = {The {Addison}-{Wesley} object technology series},
	title = {Refactoring: improving the design of existing code},
	isbn = {978-0-201-48567-7},
	shorttitle = {Refactoring},
	publisher = {Addison-Wesley},
	author = {Fowler, Martin and Beck, Kent},
	year = {1999},
}

@inproceedings{vaswani_attention_2017,
	title = {Attention is {All} you {Need}},
	volume = {30},
	url = {https://proceedings.neurips.cc/paper_files/paper/2017/file/3f5ee243547dee91fbd053c1c4a845aa-Paper.pdf},
	booktitle = {Advances in {Neural} {Information} {Processing} {Systems}},
	publisher = {Curran Associates, Inc.},
	author = {Vaswani, Ashish and Shazeer, Noam and Parmar, Niki and Uszkoreit, Jakob and Jones, Llion and Gomez, Aidan N and Kaiser, Ł ukasz and Polosukhin, Illia},
	editor = {Guyon, I. and Luxburg, U. Von and Bengio, S. and Wallach, H. and Fergus, R. and Vishwanathan, S. and Garnett, R.},
	year = {2017},
}

@article{gunawardena_concerns_2023,
	title = {Concerns identified in code review: {A} fine-grained, faceted classification},
	volume = {153},
	issn = {09505849},
	shorttitle = {Concerns identified in code review},
	url = {https://linkinghub.elsevier.com/retrieve/pii/S0950584922001653},
	doi = {10.1016/j.infsof.2022.107054},
	language = {en},
	urldate = {2024-05-25},
	journal = {Information and Software Technology},
	author = {Gunawardena, Sanuri and Tempero, Ewan and Blincoe, Kelly},
	month = jan,
	year = {2023},
	pages = {107054},
}

\end{document}